\documentclass[prc,nofootinbib]{revtex4}

\usepackage{epsfig,amsmath}

\def\epe{$e^+e^-$}

\def\Q1{{\bf Q}_1}

\def\qi{{\bf q}_i}

\def\muv{\mbox{\boldmath$\mu$}}
\def\muvs{\mbox{\boldmath${\scriptstyle{\mu}}$}}

\def\NN{\mathrm{NN}}

\def\gs{\gamma_S}

\def\ee{e$^+$e$^-$}

\def\d{{\rm d}}
\def\e{{\rm e}}
\def\t{{\rm t}}

\newcommand{\oforder}[1]{\mbox{${\cal O}(#1)$}}
\newcommand{\mc}[2]{\multicolumn{#1}{|c|}{#2}}
\def\tpm{$\pm$}
\def\tgs{$\gamma_S$ }
\def\agev{$A$ GeV}
\def\NP{\mathrm{N}_\mathrm{P}}
\def\NPt{$\mathrm{N}_\mathrm{P}$}
\def\Nbint{$\mathrm{N}_\mathrm{bin}$}
\def\fm3{$\mathrm{fm}^3$}
\def\f3{\mathrm{fm}^3}
\def\dnhm{$\d \mathrm{N}_\mathrm{h^-}/\d \eta$}
\def\be{\begin{equation}}
\def\ee{\end{equation}}
\def\L{$\Lambda$}
\def\aL{$\bar{\Lambda}$}

\begin{document}

\title{Chemical freeze-out in ultra-relativistic heavy ion collisions \\
at $\sqrt s_{NN} = 130$ and $200$ GeV.} 

\author{J. Manninen}\affiliation{INFN Sezione di Firenze, Florence, Italy}
\author{F. Becattini}\affiliation{Universit\`a di 
 Firenze and INFN Sezione di Firenze, Florence, Italy} 

\begin{abstract}
A comprehensive and detailed analysis of hadronic abundances measured in Au-Au 
collisions 
at RHIC at $\sqrt s_{\NN} = 130$ and 200 GeV is presented. The rapidity densities 
measured in the central rapidity region have been fitted to the statistical
hadronization model and the chemical freeze-out parameters determined as a 
function of centrality, using data from experiments BRAHMS, PHENIX and STAR. 
The chemical freeze-out temperature turns out to be independent of centrality 
to a few percent accuracy, whereas the 
strangeness under-saturation parameter $\gs$ decreases from almost unity in central 
collisions to a significantly lower value in peripheral collisions. Our results 
are in essential agreement with previous analyses, with the exception that fit
quality at $\sqrt s_{\NN} = 200$ GeV is not as good as previously found. 
From the comparison of the two different energies, we conclude that the 
difference in fit quality, as described by $\chi^2$ values, is owing to the
improved resolution of measurements which has probably exceeded the intrinsic
accuracy of the simplified theoretical formula used in the fits. 
\end{abstract}

\maketitle

\section{Introduction}\label{sec:introduction}

The idea of a statistical model to account for multiple hadron production processes 
in high energy collisions dates back to a work by Fermi \cite{fermi}. This model 
has been successful in reproducing the production rates of measured hadronic species 
in collisions of both elementary particles \cite{Becattini:1995if,Becattini:2001fg,
Becattini:1997rv} and heavy-ions \cite{Cleymans:1990nz,Cleymans:1992zc,
BraunMunzinger:1994xr,Becattini:1997ii,Becattini:2000jw}. These models have been 
extensively and succesfully applied to the phenomenon of multifragmentation in 
nuclear collisions \cite{multifrag}.

Many evidences have been collected that a new form of matter, the Quark--Gluon 
Plasma (QGP), where effective degrees of freedom are quarks and gluons, has 
been created in the collisions of heavy ions at relativistic energies at top SPS 
and RHIC energy. While the success of the statistical-thermal model in describing
particle multiplicities indicates that local thermodynamical equilibrium has 
been achieved and seems to confirm the general hypothesis of QGP formation, 
some aspects are still to be understood, like e.g. whether strangeness is fully 
equilibrated and the relation of this successful description with that in
elementary collisions. In this respect, an analysis of RHIC data, at a 
center-of-mass energies 130 and 
200 GeV per colliding pair of nucleon, can be illuminating. Indeed, similar 
analyses have been carried out in the past few years~\cite{Cleymans:2004pp,
Adams:2003xp,Adams:2006ke,Dumitru:2005hr,Andronic:2005yp,Baran:2003nm,Rafelski:2004dp}.
Yet, recently, a bunch of new experimental data has been published which  
makes it a worthwhile step to provide an independent analysis including this 
newly available data. Furthermore, we have studied hadron production at lower 
beam energies in heavy-ion collisions at AGS and SPS with the statistical hadronization 
model~\cite{Becattini:2003wp,Becattini:2005xt} and with this paper we complete 
our previous works. 

There is also another issue which motivates our analysis. So far, most analyses have
been using as an input to the fit $(N-1)$ particle ratios formed {\em a posteriori}
from $N$ measured particle multiplicities without including any normalizing yield.
This procedure was based on the tacit assumption that fitting either ratios or 
multiplicities lead to equivalent results. This is in general not true and the 
outcome of a statistical analysis relying on {\em only} ratios of hadron multiplicities may 
be seriously biased~\cite{Becattini:2007wt}, depending on the input set of ratios. 
The reason of this problem is that, in principle, one can form $N(N-1)/2$ different 
combinations\footnote{counting $A/B$ and $B/A$ equivalent} of particle ratios from 
$N$ different measured multiplicities and choosing a subset of them implies an 
information loss. Moreover, the different ratios are obviously correlated if a particle 
appears more than once in the ratios and those correlations must be taken into account 
in the $\chi^2$ minimization, thus complicating the fit. In the worst case, the central 
values of the fitted statistical model parameters may deviate several standard errors 
from the central values of parameters determined from a fit to particle 
multiplicities whereas the actual magnitude
of the error is not possible to know without explicit comparison. We stress that
this problem arises when using ratios calculated \emph{a posteriori} from a set
of primordially measured multiplicities, while ratios directly measured by the
experiments because of beneficial systematic error cancellation (e.g. $\pi^-/\pi^+$ or
$\bar p/p$) are perfectly safe. 
 
While at lower beam energies integrated multiplicities in full phase space are a
more suitable input for the statistical model~\cite{Becattini:2003wp}, at RHIC 
energies of $\sqrt s_{\NN} > 100$ GeV, rapidity distributions are wide enough
to allow the extraction of the thermodynamical properties of the average fireball 
produced at mid-rapidity with rapidity densities themselves. In fact, the standard width of 
charged particle rapidity distribution at $\sqrt s_{\NN} = 200$ GeV is 2.1~\cite{Bearden:2004yx}, 
sufficiently larger than standard widths of single-fireball rapidity distribution 
(at most 0.8 for pions at the kinetic freeze-out temperature of 125 MeV). On the 
other hand, at top SPS energy $\sqrt s_{\NN} = 17.2$ GeV the measured rapidity width 
is 1.3~\cite{Blume:2005kc}, which is consistently smaller and closer to the single-fireball width.
Therefore, using mid-rapidity densities at this and lower energies artificially 
enhances heavier particles with respect to lighter ones as they have a narrower
rapidity distribution.

\section{The data analysis}

We have analyzed the rapidity densities in Au-Au collisions at 130 and 200 GeV per 
participating nucleon measured by BRAHMS, PHENIX and STAR collaborations at RHIC 
employing a version of statistical hadronization model described in detail 
in~\cite{Becattini:2003wp,Becattini:2005xt}. The use of the grand-canonical formalism 
is appropriate here in that particle multiplicities are large. The formula for 
the $i$th primary hadron (including both stable hadrons and resonances) 
rapidity density reads:
\begin{equation}\label{basic}
\left< \frac {\d n_i}{\d y} \right> = \frac{\d V}{\d y} 
 \frac{(2J_i+1)}{(2\pi)^3} \int \d^3 {\rm p} \; 
\left[ \gamma_S^{-n_{si}}\e^{\sqrt{{\rm p}^2+m_i^2}/T-\muvs\cdot\qi/T} \pm 1 \right]^{-1}
\end{equation}
where $T$ is the temperature, $\qi = (Q_i,B_i,S_i)$ is a vector having as components 
the electric charge, baryon number and strangeness of the hadron and 
$\muv = (\mu_Q,\mu_B,\mu_S)$ is a vector of the corresponding 
chemical potentials; $\gamma_S$ is the strangeness under-saturation factor and
$n_{si}$ is the number of valence strange quarks in the $i$th hadron; the upper 
sign applies to fermions, the lower to bosons.
The absolute normalization $\d V/\d y$ in (\ref{basic}) is a product of the
rapidity density of clusters at mid-rapidity $\rho(0)$ times the volume of the average fireball 
at mid-rapidity \cite{Becattini:2005xt}. For the above formula to make sense, the 
parameters $T$, $\muv$ and strangeness under-saturation factor $\gamma_S$ should 
be constant over a rapidity range encompassing the single fireball rapidity width
\cite{Becattini:2005xt}. In this work, the chemical potential $\mu_S$ is determined 
enforcing vanishing strangeness density and $\mu_Q$ by requiring the final ratio 
of charge to baryon number to equalize the initial one $Z/A$, i.e. by assuming
that there is no major dependence of the strangeness and electric density on
rapidity. The other 4 parameters ($T,\mu_B,\gs,\d V/\d y$) are determined by 
minimizing the $\chi^2$:
\begin{equation}
\chi^2=\sum_i \frac{(\d N_i^\e/\d y - \d N_i^\t /\d y)^2}{\sigma_i^2}
= \sum_i \frac{(\d N_i^\e/\d y - n_i^\t \, \d V /\d y)^2}{\sigma_i^2} 
\end{equation}
in which $\d N^e_i/\d y$ and $\d N^t_i/\d y$ are the experimental and theoretical 
rapidity densities, $n_i^t$ is the particle density evaluated within statistical 
hadronization model and $\sigma_i$ is the experimental error of the rapidity
density of a particle species $i$. Unless otherwise stated, all experimental errors
quoted in this paper are a quadratic sum of statistical and systematic errors.
Before going deeper into the data analysis and exploration of results, we need to 
discuss a preliminary treatment of the experimental data which was necessary to
make a combined analysis of all experiments.

\subsection{Centrality interpolations}\label{sec:interpolation}

Particle rapidity densities are measured in some selected centrality windows which 
are different for different experiments. As a consequence, measurements relevant 
to e.g. the most central collisions from different experiments must be renormalized. 
Moreover, the chosen centrality windows can be different for different particle 
species even within the same experiment. Therefore, in order to make a correct 
analysis of the full data set, one needs to find a proper method to estimate rapidity 
densities of different hadronic species {\em in the same centrality} window,
i.e. a proper interpolation method.
This can be done in many ways. For example, taking advantage of a possible linear 
and parabolic scaling with number of participants (\NPt) or number of binary collisions
(\Nbint) as well as with the negative hadron pseudo-rapidity density (\dnhm). In many 
cases such a simple scaling behaviour can be found. For example, STAR collaboration has 
found out~\cite{Adler:2002uv} that \L~and \aL~rapidity densities scale well with the 
\dnhm~in Au-Au collisions at 130\agev. However, none of the above-mentioned scaling variables with 
simple functional forms is able to describe the centrality dependence of all different 
hadron species; instead, more complex dynamical combination (e.g.~\cite{Eskola:1999fc,
Kharzeev:2000ph}) of ``hard'' and ``soft'' physics processes must be considered. 

Typically the interpolation correction is not very large. For instance, we 
might want to estimate the rapidity density of a hadron species in the [0-6\%] most 
central collisions while the experimental value is given for the [0-5\%] and [5-10\%] 
most central collisions. It is also important to note that centrality fractions are 
independent of the observable used to define them provided that the observable 
varies monotonically; this ensures the one-to-one correspondence between different
observables. We have chosen an interpolation method we deem is more robust and
model-independent than any simple scaling with \NPt, \Nbint~or \dnhm~and 
implemented it consistently for all\footnote{there are few exceptions to this rule 
which will be explicitly discussed} rapidity densities that need to be interpolated. 
We write the rapidity density of a hadron species $i$ as a $k_\mathrm{max}$th order 
polynomial of the centrality as follows:
\be\label{3rdor}
\frac{\d^2 N_i}{\d y \d c} = \sum_{k=0}^{k_\mathrm{max}} \alpha_k^i c^k \qquad ; 
 \qquad c = 1 - x
\ee
in which $\alpha_k^i$ denote free parameters and $x\in[0,1]$ is the fraction of the
differential cross section as a function of the variable defining centrality itself 
(0=0\% most central and 1=100\%  most central collisions). The rapidity density 
of a hadron species in a certain centrality window $[y_\mathrm{min}, y_\mathrm{max}]$ 
is obtained by integrating Eq. (\ref{3rdor}) over $[y_\mathrm{min}, y_\mathrm{max}]$. 
We thereby calculate the rapidity densities in the centrality windows where they 
have been measured and fit the $k_\mathrm{max}$+1 free parameters to reproduce the 
measured rapidity densities. Once the parameters are fitted, rapidity densities 
of any hadron species can be estimated in any centrality window 
$[y'_\mathrm{min},y'_\mathrm{max}]$ by simply integrating with respect to $c$ over 
the desired region of centrality keeping the fitted parameters fixed. 
The maximal order of the polynomial (i.e. maximal number of free parameters minus one) 
for a certain particle species $i$ is the number of centrality bins in which it is 
measured. We have always chosen the maximal order $k_\mathrm{max}$=$N_{windows}$-1 
for these interpolations unless\footnote{this happened in one case} 
this would lead to non-monotonic behaviour of Eq. (\ref{3rdor}) within the centrality 
range we are interested in, in which case we have chosen the maximal order such 
that the polynomial is monotonically increasing. The original experimental errors 
are properly propagated so that the errors of our interpolated rapidity densities 
include the original errors as well as the additional uncertainty arising from the 
interpolation. 

\begin{figure}[!t]
$\begin{array}{c@{\hspace{-0.05cm}}c}
\epsffile{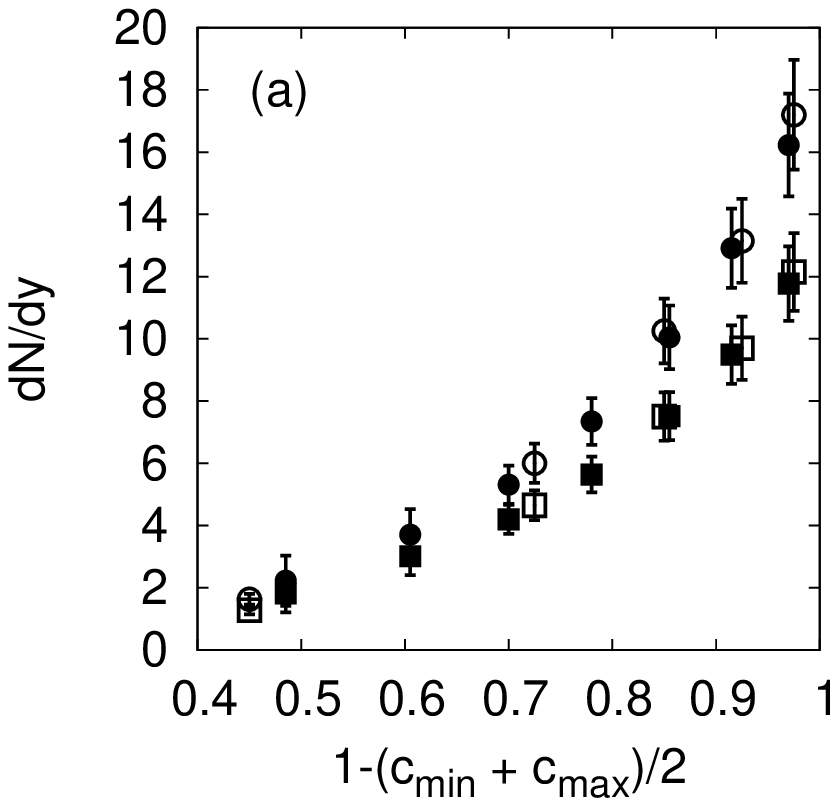} & 
\epsffile{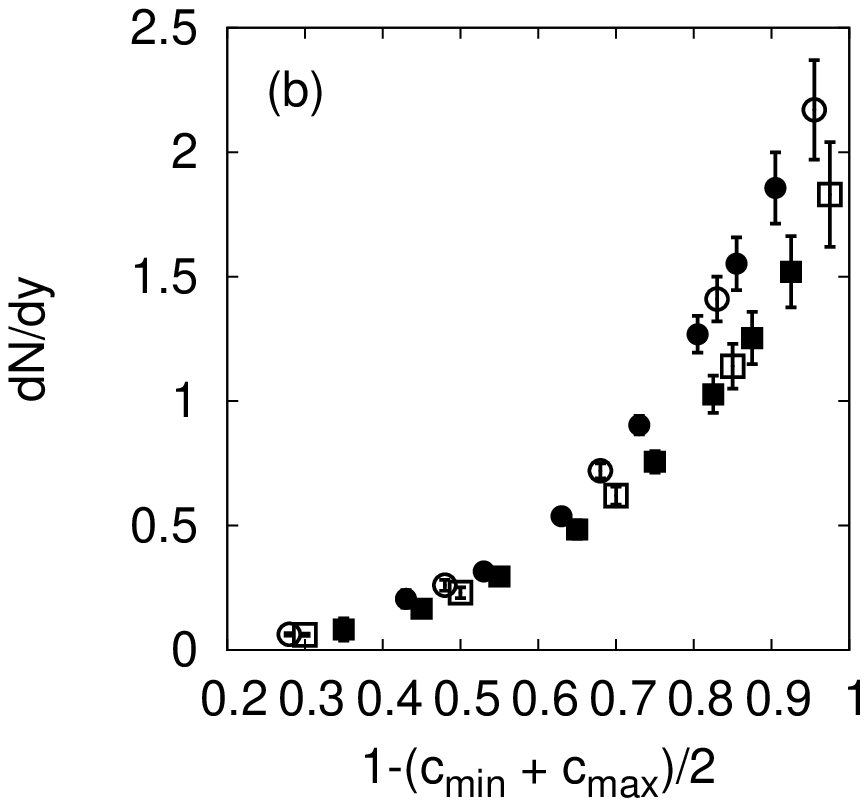}
\end{array}$
\caption{Left panel (a): Experimental \L~(open round symbols) and \aL~(open square 
symbols) rapidity densities as a function of centrality 
($c_\mathrm{min}$ and $c_\mathrm{max}$ being the borders of the centrality window 
under consideration)
in Au-Au collisions at 130\agev~measured by STAR collaboration~\cite{Adler:2002uv}. 
Full symbols denote our interpolated rapidity 
densities in centrality windows the statistical analysis is carried out.\\
Right panel (b): Measured $\Xi^-$ (open round symbols) and $\Xi^+$ (open square symbols) 
rapidity densities in Au-Au collisions at 200\agev~\cite{Adams:2006ke}. 
The full symbols denote our estimates for the same rapidity densities in the 
centrality windows the statistical analysis is carried out. The round symbols 
are shifted 0.02 leftward for clarity.\\
$c_min$ and $c_max$ denote the limits of the centrality bin corresponding to 
each data point; in other words, the data points have been stick in the centre 
of the bin. 
\label{fig:interpolation}}
\end{figure}

An example of our interpolations is shown in fig.~\ref{fig:interpolation}. In the 
left panel, \L~(open round symbols) and \aL~(open square symbols) rapidity densities 
measured by STAR collaboration at 130\agev~are shown while the filled symbols denote 
our estimates for the corresponding rapidity densities in 7 other centrality windows, 
namely the ones in which kaons and nucleons are measured. In the right panel of fig.~
\ref{fig:interpolation}, similar plot is shown for $\Xi^-$ and $\Xi^+$ at 200\agev. 

The experimental rapidity densities as well as our interpolated values are shown in 
tables \ref{input130}, \ref{input200_1} and \ref{input200_2}. The numbers in plain
text are our estimates while numbers written in bold case are the experimental values
our interpolations are based on. 

\subsection{Au-Au collisions at 130\agev}

STAR collaboration has measured $p$~\cite{Adams:2003ve},
$\bar{p}$~\cite{Adams:2003ve}, $K^+$~\cite{Adler:2002wn} and 
$K^-$~\cite{Adler:2002wn} rapidity densities around mid-rapidity in Au-Au
collisions at 130\agev~in 8 centrality bins, see table \ref{input130}. Proton and 
anti-proton rapidity densities include weak feeding from (multi)strange hyperons.
We have chosen to perform our statistical model analysis as a function of centrality 
at this beam energy in the centrality windows nucleons and kaons are measured.

On the other hand, $\Lambda$ and $\bar{\Lambda}$ rapidity densities are measured in 5 
different centrality bins~\cite{Adler:2002uv} and the rapidity densities also include 
weak feeding from multi-strange hyperons. By fitting different functions to the transverse 
mass spectra, STAR collaboration has obtained two slightly different values for both 
$\Lambda$ and $\bar{\Lambda}$ rapidity densities, thus we have taken a weighted 
average of those. Since $\Lambda$ and $\bar{\Lambda}$'s centrality bins differ 
somewhat from the centrality bins in which nucleons and kaons are measured, we have 
estimated the hyperon yields in the latter ones as described in the previous section. 

For the fit to be reliable, it is necessary to include the most abundantly produced
particles, charged pions, in the analysis. However, the integrated STAR 
$\pi^\pm$ rapidity densities (corrected for weak decay feeding) are publicly available 
only in the 0-5\% most central bin at 130\agev~\cite{Adams:2003yh} and so polynomial 
interpolation for peripheral collisions can not be implemented. Without better knowledge, 
we have assumed that the ratio 
\be\label{pis}
\frac{\d N_{\pi^+}}{\d y} \Big/ \frac{\d N_{h^-}}{\d \eta}  
= \frac{\d N_{\pi^-}}{\d y}\Big/ \frac{\d N_{h^-}}{\d \eta} 
= 239.0 / 296.6
\ee
retains its value in every centrality bin. Since most of the charged hadrons 
emitted in a heavy-ion collision are pions, we deem that Eq.~(\ref{pis}) yields 
a reasonable estimate of the pion rapidity densities. As far as the error on the
estimated rapidity densities is concerned, we have have added the additional systematic 
error arising from the extrapolation to different centrality bins in quadrature with 
the relative error of 4.5\% that is quoted in the 0-5\% most central collisions. 
Based on the published~\cite{Adler:2002wn,Adler:2002uv} systematic errors of \dnhm~in 
the various centrality bins, we estimate an overall error of 10.3\% in the pion rapidity 
density. In order to check the stability of our analysis, we have repeated the fits
by assuming 5\% and 15\% error in the pion rapidity densities in each centrality bin. 
The ensuing fitted parameters showed little difference and their central values
turned out to be well within the error bar of the main fit.

The rapidity densities of hyperons $\Xi^-$ and $\Xi^+$ have been measured 
in 3 centrality windows (0-10\%, 10-25\% and 25-75\% most central collisions) while 
$\Omega + \bar{\Omega}$ is measured in the 0-10\% most central collisions~\cite{Adams:2003fy}
only. Similarly to \L~and \aL~, two slightly different rapidity densities are quoted for 
both $\Xi$'s and we have taken the weighted average as our input for the analysis. 
Finally, $\phi$-meson is measured in 3 different centrality windows (0-11\%, 11-26\% 
and 26-85\% most central collisions)~\cite{Adler:2002xv}. Similarly to $\Lambda$'s, 
we have estimated the $\Xi^\pm$ and $\phi$ rapidity densities in the 8 STAR reference
centrality bins, but since data is available in 3 centrality windows only, a 2nd order 
interpolation polynomial was used in Eq.~(\ref{3rdor}). 

Because of the vast width of the peripheral centrality window ($\approx 30-80$\% most 
central collisions), our method to estimate $\Xi$ and $\phi$ rapidity densities fails 
in the most peripheral bin (58-85\%), and in general the relative errors increase 
with decreasing centrality. Particularly, the extrapolation of $\Omega+\bar{\Omega}$ 
rapidity density from central to peripheral collisions based on a single centrality 
is meaningless. Thus, we have removed $\Omega$ from the STAR particle set to estimate 
the freeze-out parameters in Au-Au collisions at 130\agev~in the 7 most central bins. 
As a check, the analysis has been repeated in the two most central bins by including 
$\Omega+\bar{\Omega}$ rapidity density, which we have assumed to scale with the negative
hadron pseudo-rapidity density, at least in this short range. The fit outcome turns
out to be essentially unaffected by this inclusion and thus all quoted results in 
this work at 130\agev~refer to fits without $\Omega$'s.  

Also PHENIX collaboration has measured $\pi^+$, $\pi^-$, $K^+$, $K^-$, p and $\bar{p}$ 
rapidity densities in a pseudo-rapidity window of $|\eta|<0.35$ around mid-rapidity 
in Au-Au collisions at 130\agev~\cite{Adcox:2001mf,Adcox:2003nr}. The data is divided 
in 5 centrality bins (see table \ref{input130}) which differ from STAR's ones and thus 
direct comparison is not possible. Also, $\Lambda$ and $\bar{\Lambda}$ rapidity densities 
have been measured in the most central bin at 130\agev~\cite{Adcox:2002au}. No weak 
decay corrections were applied to any of the hadron species. We have repeated the fits 
with the PHENIX data and found out that the data set is rich enough to fix all the 
statistical model free parameters only in the most central bin, in which hyperons are
included in the data sample (see fig.~\ref{fig:rhic130_1} and table \ref{parameters130}). 
In the other bins, the set $\pi^\pm$, $K^\pm$, p and $\bar{p}$ does not allow to 
reliably fit all the 4 free parameters because of the relatively short lever arm in 
mass and the low baryon chemical potential which makes pions multiplicities too close.

Finally, we have made a combined fit to PHENIX and STAR data. First, the PHENIX 
rapidity densities of $\pi^\pm$, $K^\pm$, p and $\bar{p}$ have been estimated in the 
STAR centrality bins (see again table \ref{input130}) according to the aforementioned
interpolation procedure. The obtained $K^\pm$ and nucleon rapidity densities agree 
very well with the corresponding experimental STAR values in the most central bin
while the relative discrepancy increases in the more peripheral ones, yet within 
the error bars. It should also be pointed out that PHENIX rapidity densities of pions 
are larger than the corresponding STAR values because in the former case no weak 
decay corrections was applied. The possible different overall normalization between 
the 2 experiments was taken into account by introducing one more free parameter $f_P$ 
multiplying all PHENIX rapidity densities; otherwise stated, the common scaling 
factor $dV/dy$ becomes $f_P \times dV/dy$ for PHENIX data. Obviously, one expects 
$f_P\approx$ 1 in each centrality bin, if the experiments are to be in essential
agreement. 

The statistical model best fit parameters determined from the combined STAR+PHENIX 
fit are shown in fig.~\ref{fig:rhic130_1} and table \ref{parameters130} along
with those determined by a fit to STAR data alone. It can be seen that they are
in very good agreement with each other and that the cross-normalization parameter
$f_P$ varies between 0.8 to 0.92 throughout the examined centrality range.

\subsection{Au-Au collisions at 200\agev}

The analysis has been carried out also at $\sqrt s_{\NN} = 200$ GeV. At this energy, 
PHENIX collaboration has measured~\cite{Adler:2003cb} the rapidity densities of the 
same set of hadrons as at $\sqrt s_{\NN} = 130$ GeV, i.e. $\pi^+$, $\pi^-$, $K^+$, 
$K^-$, $p$ and $\bar{p}$ over 11 centrality bins. Pion rapidity densities do not 
include weak decay products while proton and anti-proton rapidity densities are 
corrected from $\Lambda$ and $\bar{\Lambda}$ feeding. Like at the lower beam 
energy, the set of different hadron species is not large enough for us to fit the 
statistical model parameters reliably. 

STAR collaboration has measured the same hadron species ($\pi^\pm$, $K^\pm$, $p$, 
$\bar{p}$, $\Lambda$, $\bar{\Lambda}$, $\phi$, $\Xi^\mp$ and $\Omega+\bar{\Omega}$
~\cite{Adams:2003xp,Adams:2006ke,Abelev:2007rw,Chen:2008ix}) as at $\sqrt s_{\NN} = 130$ GeV. The 
centrality windows in which STAR and PHENIX have measured pions, kaons and protons
are mostly the same. In order to minimize the amount of data manipulation, we have 
chosen to keep the PHENIX data as it is and to estimate STAR rapidity densities,
whenever necessary, in PHENIX centrality bins. The hyperons and $\phi$ meson have 
been measured in many more centrality windows than at the lower beam energy. 
Still, most of the windows are wider than the ones for pions and so we have interpolated hyperon 
rapidity densities in the narrower pions centrality windows according to Eq. (\ref{3rdor}). 
All rapidity densities measured by STAR collaboration are cleaned from weak decay 
products except $p$ and $\bar{p}$ which include feeding from hyperons.

Finally, BRAHMS collaboration has measured the same hadron rapidity 
densities~\cite{Arsene:2005mr} as PHENIX collaboration in 4 different centrality 
windows. The pion rapidity densities do not include any weak decay products while 
only \L~and \aL~decay products are subtracted from nucleon rapidity densities. We 
have estimated the BRAHMS rapidity densities in the same centrality bins defined by
PHENIX collaboration. 

Similarly to the lower beam energy, we have determined the chemical freeze-out 
parameters by performing a fit to STAR data alone and then a combined fit to STAR, 
PHENIX and BRAHMS rapidity densities. In the combined fit, free parameters $f_P$ and 
$f_B$ multiplying theoretical rapidity densities of PHENIX and BRAHMS respectively 
have been introduced in order to take into account possible discrepancy in overall 
normalization among different experiments. As it seems that there is a significant 
discrepancy in the $\Lambda/p$ among the three experiments, we have decided to 
exclude proton and anti-proton rapidity densities measured by PHENIX and BRAHMS 
collaborations in the analysis (see detailed discussion in Section~\ref{discussion}).
The resulting statistical model best fit parameters are shown in 
table~\ref{parameters200} and fig.~\ref{fig:rhic200_1}.

\subsection{Further notes}

We have left out from our analysis some additional rapidity densities of hadron 
species that are measured at RHIC. The PHENIX measurement~\cite{Adler:2004hv} of 
$\phi$ meson at 200\agev~is left out from our analysis due to the very large 
discrepancy with the corresponding STAR values. We have compared the statistical 
model predictions for $\phi$ meson production with the PHENIX measurement though, 
and found out a severe disagreement between the statistical model prediction and 
the PHENIX measurement (of the same order as between STAR and PHENIX measurements).

Also, STAR measurements~\cite{Adams:2004ep,Adams:2006yu} of strange resonances 
$K(892)$, $\Sigma(1385)$ and $\Lambda(1520)$ are left out from our analysis due 
to their very short lifetime which makes their decay products rescatter after 
chemical freeze-out, a known issue (see e.g.~\cite{Becattini:2003wp}) in statistical 
model analysis \emph{in heavy-ion collisions}. On top of above measurements, we 
have omitted the STAR $K^0_S$~\cite{Adler:2002wn} measurement in our analysis. 
Within the statistical hadronization model, $K^0_S$ multiplicity is always
between the $K^+$ and $K^-$ yields while the STAR measurement suggests much 
smaller $K^0_S$ multiplicity compared with both $K^+$ and $K^-$. Thus, in this 
case statistical hadronization model would not be able to reproduce all $K$ 
rapidity densities on a satisfactory level and so we have decided to rely on the 
$K^\pm$ yields only.
In order to take into account the additional uncertainty on parameters implied 
in fits with $\chi^2/dof>1$, parameter errors have been rescaled by 
$\sqrt{\chi^2/dof}$ if this is larger than 1, according to Particle Data Group 
procedure \cite{pdg}.

\section{Discussion}\label{discussion}

\begin{figure}[!t]
$\begin{array}{c@{\hspace{-0.05cm}}c}
        \epsffile{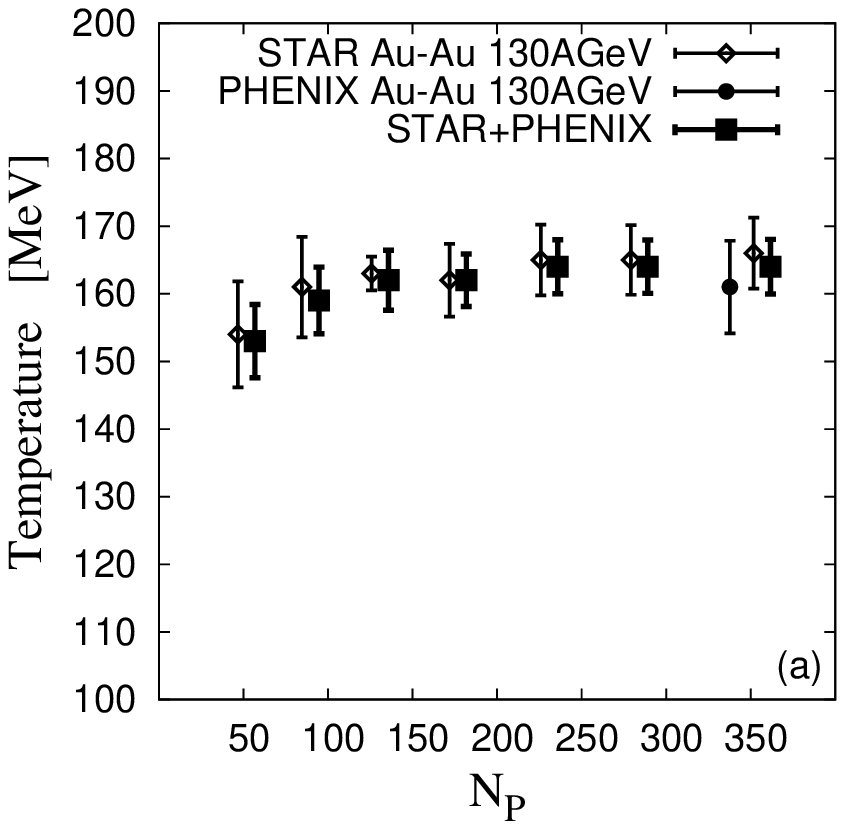} & \epsffile{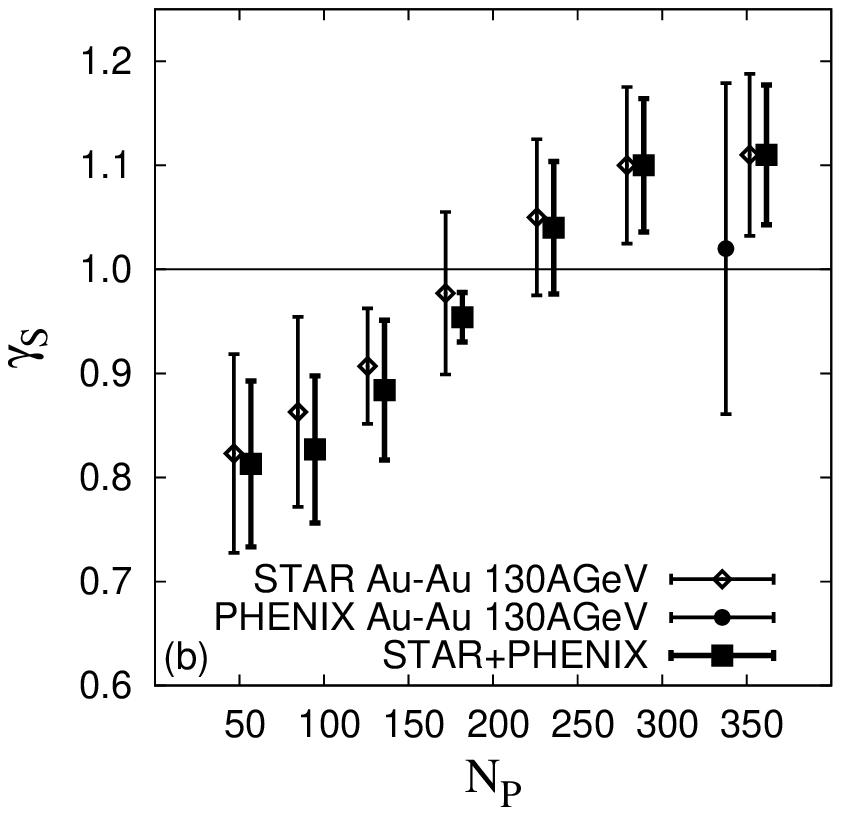} \\
        \epsffile{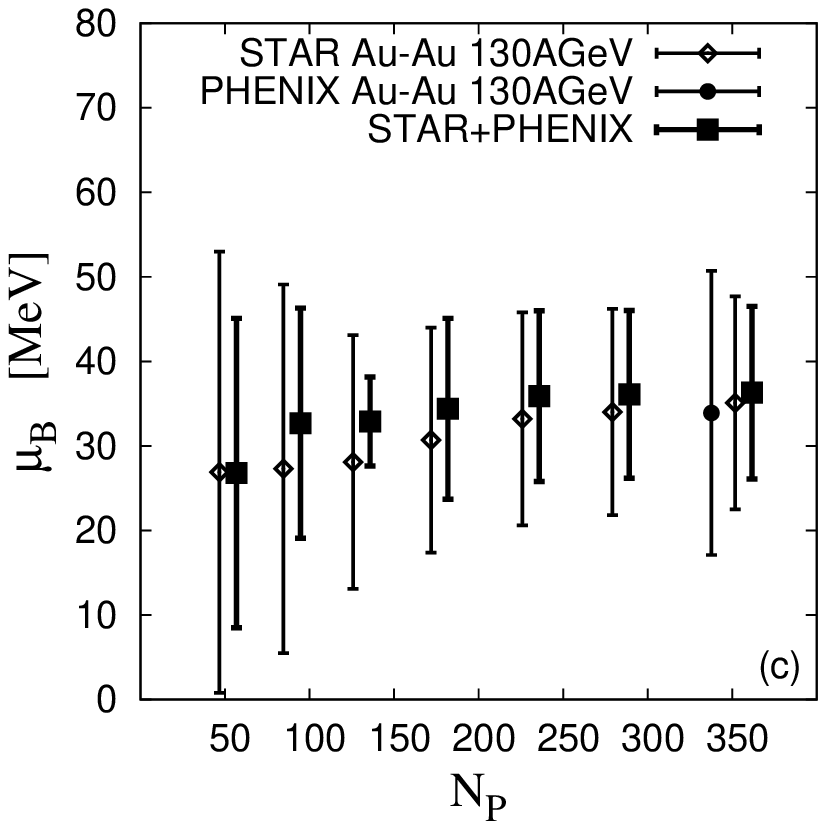} & \epsffile{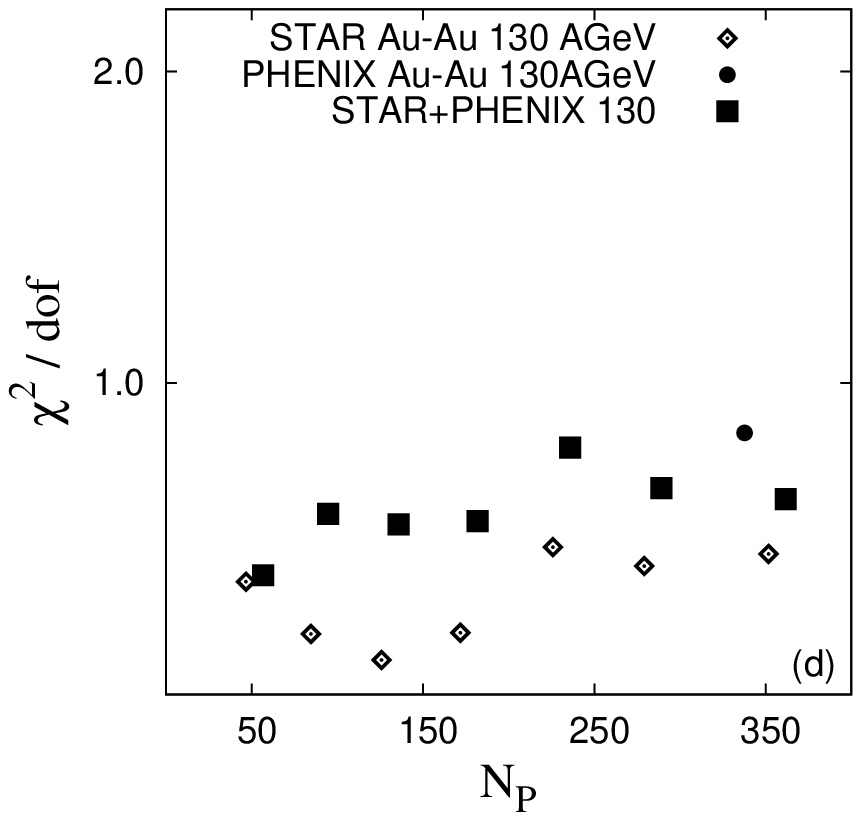} \\
\end{array}$
\caption{Chemical freeze-out temperature (top left panel (a)), 
strangeness under-saturation parameter \tgs~(top right panel (b)), 
baryon chemical potential (bottom left panel (c)) and 
the best fit $\chi^2$ per degrees of freedom (bottom right panel (d))
as a function of collision centrality in Au-Au collisions around mid-rapidity at 
130\agev. Open symbols represent fits to STAR data while square full symbols 
represent fits to combined STAR+PHENIX data and the full round symbol a fit to 
PHENIX data alone. The full square symbols are shifted 10 units rightward and 
the full round symbol 10 units leftward for clarity.\label{fig:rhic130_1}}
\end{figure}

\begin{figure}[!t]
$\begin{array}{c@{\hspace{-0.05cm}}c}
        \epsffile{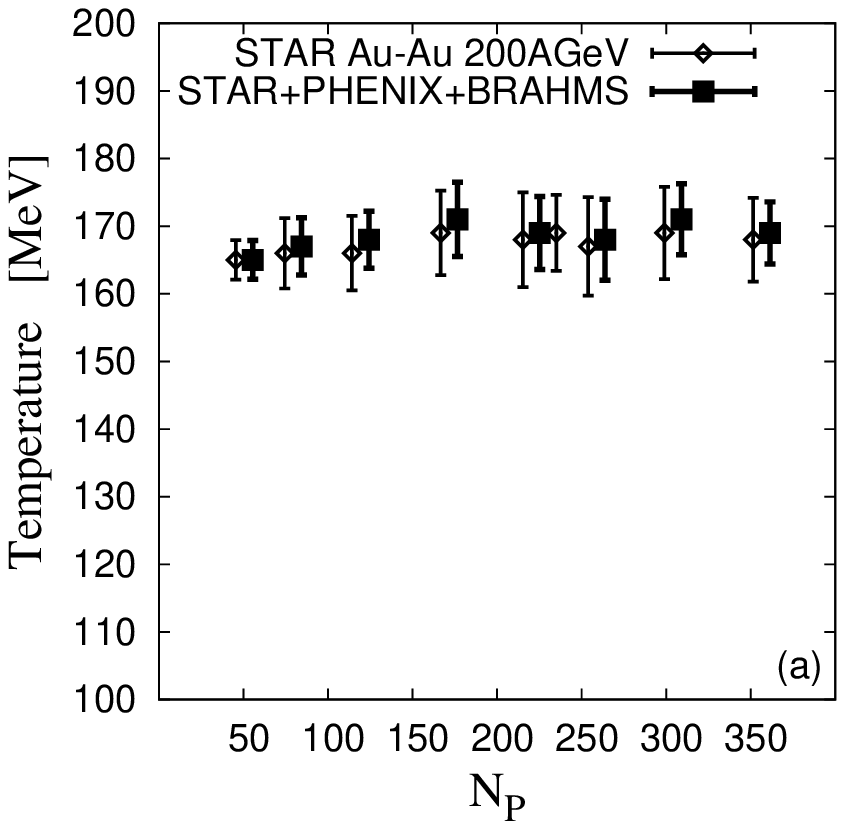} & \epsffile{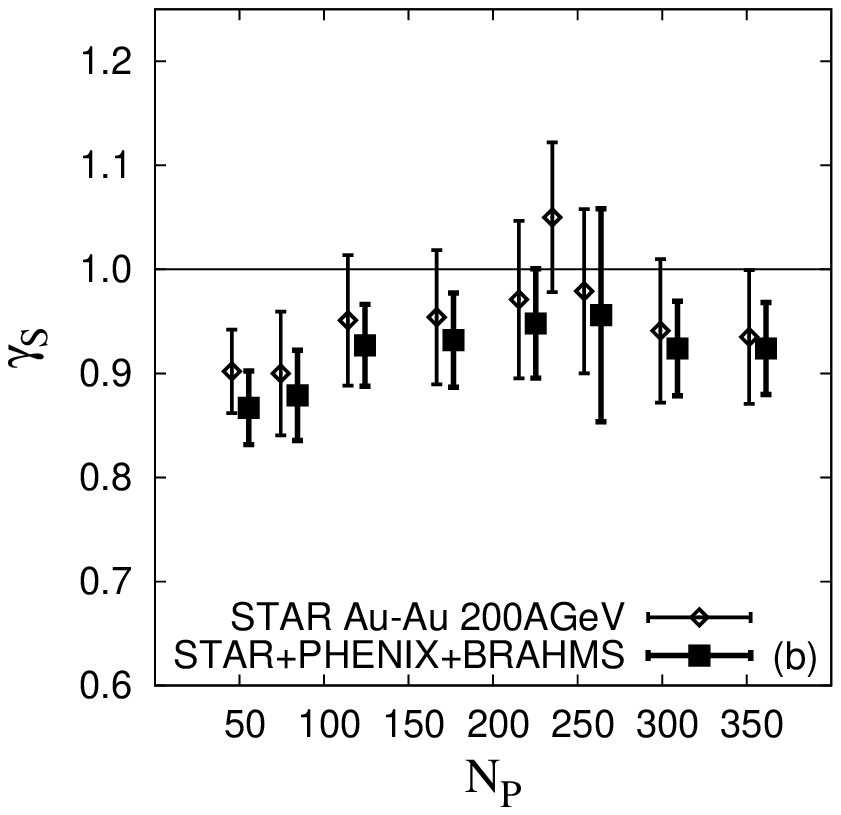} \\
        \epsffile{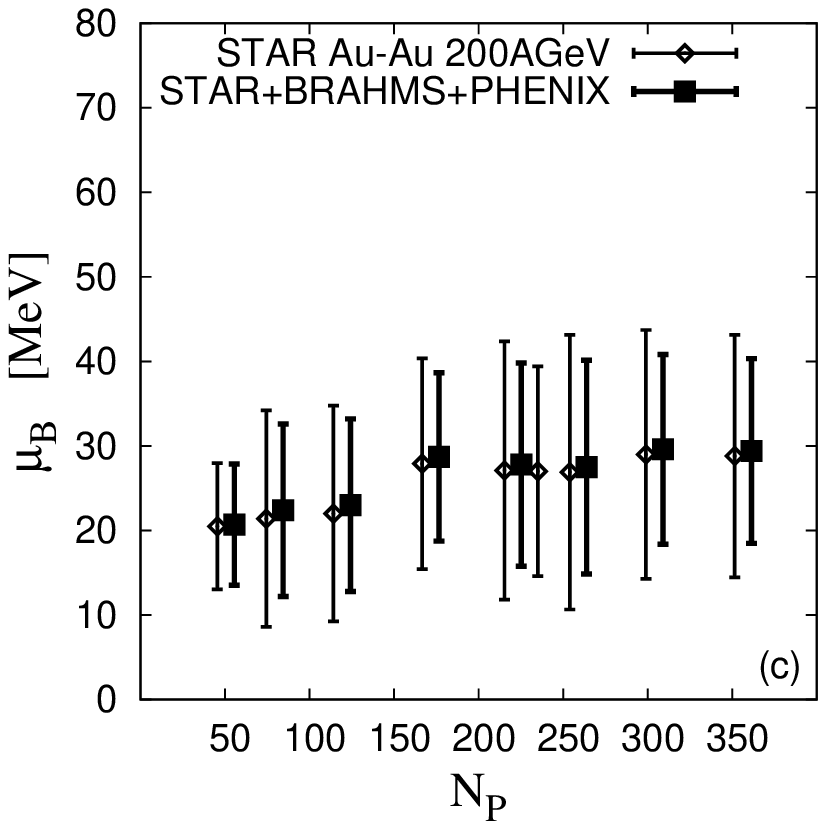} & \epsffile{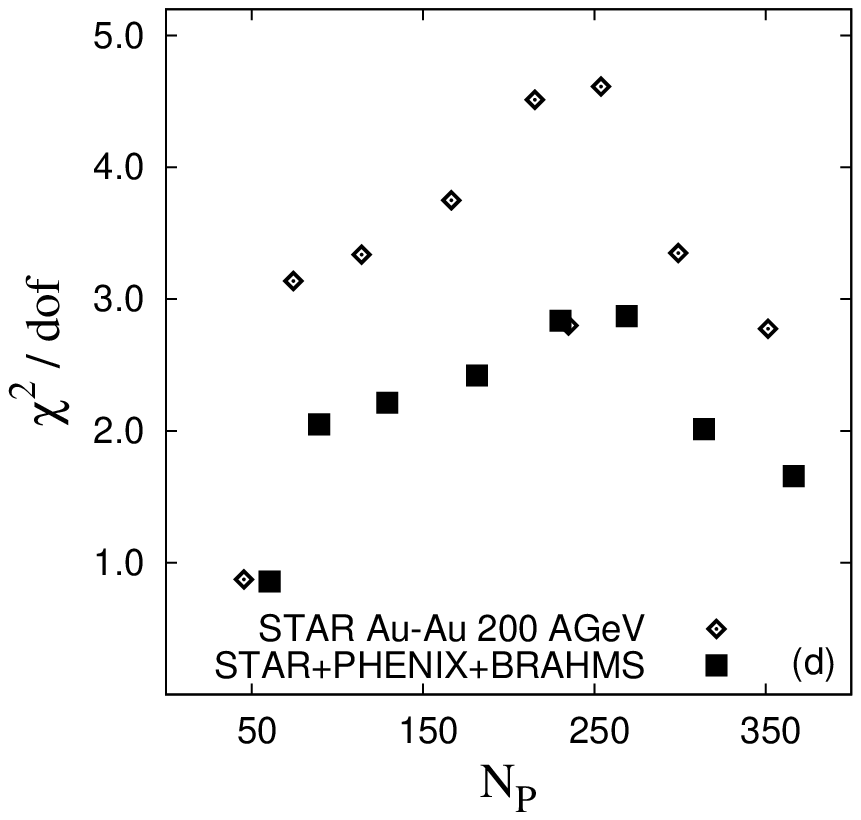} \\
\end{array}$
\caption{
Chemical freeze-out temperature (top left panel (a)), strangeness under-saturation 
parameter \tgs~(top right panel (b)), baryon chemical potential (bottom left 
panel (c)) and the best fit $\chi^2$ per degrees of freedom (bottom right panel (d))
as a function of collision centrality in Au-Au collisions around mid-rapidity at 200\agev. 
Full symbols represent fits to combined STAR+PHENIX+BRAHMS data while the 
open symbols represent fits to STAR data only. 
The filled symbols are shifted 10 units rightward for clarity.\label{fig:rhic200_1}}
\end{figure}

Looking at the figs.~\ref{fig:rhic130_1} and \ref{fig:rhic200_1}, the most 
striking feature of statistical hadronization model fits is that temperature and 
baryon-chemical potential do not show much dependence on centrality. Particularly, 
temperature is constant at few percent level. The strangeness phase-space under-saturation
parameter $\gamma_S$ seems to be somewhat smaller than unity in peripheral collisions 
but reaches unity in semi-central collisions and then saturates. All of this
is in agreement with previous findings~\cite{Cleymans:2004pp,Adams:2006ke}.
Indeed, the increasing trend of $\gamma_S$ as a function of centrality is more 
evident at $130$ GeV than at 200 GeV; furthermore, at 130 GeV $\gamma_S$ apparently 
exceeds 1 in the most central collisions. However, 
given the large error bar, this parameter is still consistent with its natural 
saturation value, i.e. 1. With the present level of accuracy, we believe that no 
claim can be made about different values of $\gamma_S$ at the two energies. 

Comparing the statistical hadronization model parameters among the two 
different beam energies with the same $\NP$, we see very little differences. 
The resulting chemical freeze-out temperatures, \tgs factors and scaling volumes are 
very similar and we can see mild beam energy dependence in the baryon chemical 
potential only. Thus, it seems that at RHIC energies we have reached a 
saturation limit in which hadrons decouple from the strongly interacting system 
at mid-rapidity in almost the same thermodynamical state. We then easily predict,
in agreement with others, that Pb-Pb collisions at the LHC will find $T\approx170$ MeV
and $\gamma_S\approx1$.

The statistical hadronization model describes the STAR data very well at 
130\agev~in every centrality bin. The $\chi^2/dof$ is less than 1 (see table~
\ref{parameters130} and fig.~\ref{fig:rhic130_1}) both with STAR data alone as 
well as with the combined STAR + PHENIX data.
The resulting cross normalization factors $f_P$ in the combined fit are around
0.9 in the central and semi central collisions while in the peripheral systems we
find rather low factors of the order of 0.8. The same tendency is already visible
when comparing the experimental STAR rapidity densities and our interpolated
PHENIX rapidity densities at 130\agev. Up to what extent this is a manifestation 
of a true difference in absolute normalization among these two experiments in 
peripheral collisions or a fit artefact is not possible to decide based on the 
published data available because there is no overlap in the particle sample.
At this energy, it seems that statistical hadronization model tends to 
overestimate the proton and anti-proton rapidity densities while other particle 
species are very well described and no systematic discrepancy is seen between 
data and model. 

Conversely, at $\sqrt s_{\NN}=200$ GeV, the $\chi^2/dof$ values (see 
table~\ref{parameters200} and fig.~\ref{fig:rhic200_1}), 
are larger than one. This is simply due to the better accuracy of measurements at
the larger beam energy while the average relative deviations of calculated rapidity 
densities from the experimental ones are of the same order at both energies (in fact, 
they are 9.7\% and 11.5\% in central heavy-ion collisions at 130\agev~and 
200\agev~respectively). 
The residuals (defined as the ratio between the difference model-data and the 
experimental error) and relative deviations of measured and calculated rapidity 
densities in central Au-Au collisions at 130 and 200\agev~are shown in tables 
\ref{combo130_ff1} and \ref{combo200_f1} where one can see that the model is able 
to reproduce the data at the same level of accuracy at both beam energies. 

For this reason, it should be stressed that the $\chi^2$ test should be used 
very carefully in order to avoid naive and hasty judgements about the validity 
of the model. Indeed, what these values at different energies tell us is that the 
simple formula (\ref{basic}) is valid up to some level of resolution and fails when 
the accuracy of measurements exceeds it, i.e. at 200\agev. This is really no surprise 
because formula (\ref{basic}) relies on several side-assumptions and approximations 
that are not expected to be exactly fulfilled. In other words, the theoretical 
model expressed by (\ref{basic}) is to be taken as a zero-order approximation and 
not as a precise representation of the real process. When the resolving power 
of experiments is good enough, higher order corrections become necessary, 
although they are very difficult to estimate and implement. For instance, an
assumption which may not be exactly true is the vanishing of strangeness density 
at mid-rapidity, which has been used in our fits; clearly, treating $\mu_S$ as 
a further free parameter could reduce the $\chi^2$. Another important approximation
is concerned with the hadron-resonance gas model, where both hadrons and 
resonances are handled as free particles with distributed mass and the 
contribution of non-resonating interactions among stable hadrons is neglected;
it is clear that this approximation will fail at some very good resolution. 
Finally, it should be reminded that the sharp separation between chemical and 
kinetic freeze-out is also an idealization. Even though hadronic rescattering 
does not play a major role in determining particle abundances (one good evidence
is the success of the statistical model itself), we know that it is there; thus,
different inelastic reactions may cease at different stages of the post-hadronization 
expansion and this involves deviations from the simple scheme of elastic-inelastic 
separation.

Nevertheless, the $\chi^2$ fit is a useful tool to determine the best parameters 
of the zero-order theory but should be used with care as an absolute measure of 
the fit quality. For example, the relative errors of hadron multiplicities in 
\epe~experiments are typically few percent only, which leads to relatively large $\chi^2/dof$ 
values~\cite{Becattini:2008tx} at least when compared with the $\chi^2/dof$ 
values in heavy-ion collisions in which the relative errors of multiplicities 
are typically larger and thus a blind comparison of the $\chi^2/dof$ values arising 
from the fits to elementary collisions and heavy-ion collisions could be highly
misleading. However, as has been mentioned, if fits have a low quality, the
estimated parameter errors could be unrealistically small and this is why 
we rescaled errors by $\sqrt{\chi^2/dof}$, according to the procedure adopted
by Particle Data Group in such cases \cite{pdg}.

\begin{figure}[!h]
\epsfxsize=4.5in
        \epsffile{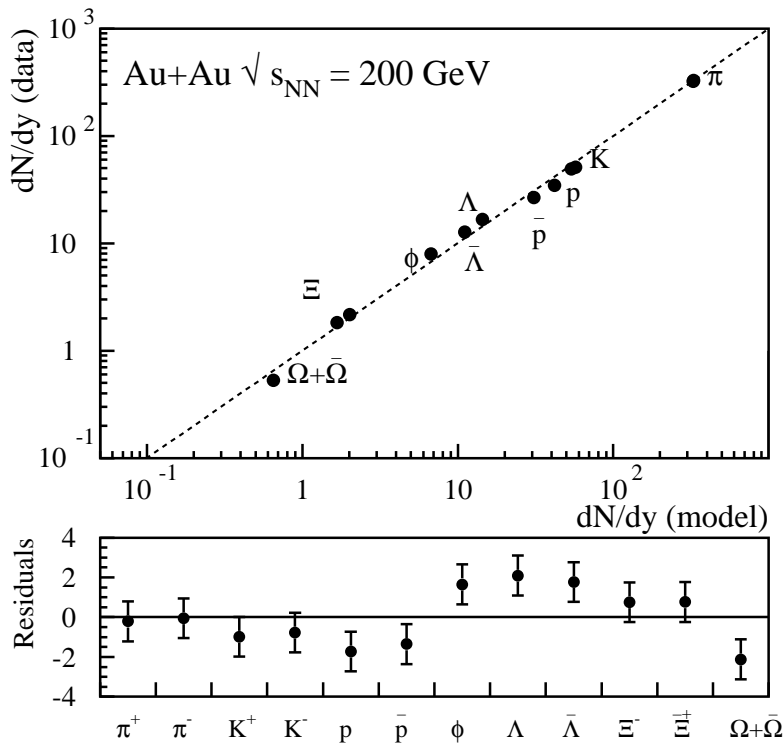}
\caption{Above: measured versus fitted rapidity densities in Au-Au collisions 
at $\sqrt s_{NN}=200$ GeV. Below: residual (defined here as the ratio between 
the difference data-model and the experimental error) distribution. The data 
is from STAR, the model values refer to the combined fit to BRAHMS, STAR, PHENIX 
experiments.
\label{rhic200}}
\end{figure}

Fits to STAR data at 200\agev~do not reveal any particular discrepancy between 
the experiment and the model, all particle species are roughly equally well 
described ($\sim 1\sigma-2\sigma$ deviation) with the exception of pions which 
are always very well reproduced (see fig.~\ref{rhic200}). We have repeated the 
fits by systematically 
removing different particle species from the fits and found out that similarly 
to the lower beam energy, statistical model seems to have problems in reproducing 
the proton and anti-proton rapidity densities especially together with the 
$\Lambda$ and $\bar{\Lambda}$ rapidity densities. Removal of either of these two 
particle species (and their antiparticles) will lead to much smaller $\chi^2/dof$ 
values while the resulting best fit parameters are adjusted within the errors
of the parameters resulting from a fit to the full data set. In general, the 
statistical hadronization model tends to systematically over estimate the 
$\Omega+\bar{\Omega}$ rapidity density and under estimate the other hyperon yields 
at 200\agev. 

\begin{figure}[!ht]
        \epsffile{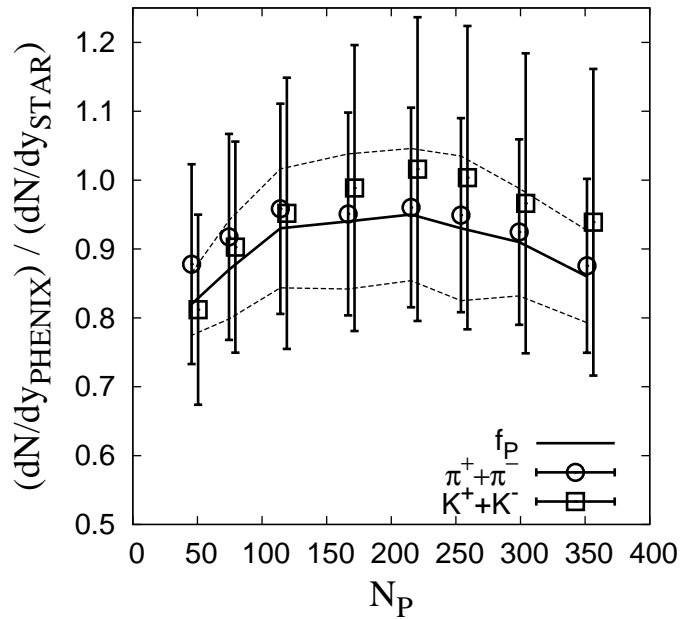}
\caption{The average rapidity density of pions (open circles) and 
kaons (open squares) measured by PHENIX collaboration divided by the same 
quantity measured by STAR in Au-Au collisions at 200\agev~as a function of 
centrality. The fitted cross-normalization factor $f_P$ (thick line) closely 
follows the measured ratio of pions. Dashed lines visualize the errorband of $f_P$.
\label{picrhic}}
\end{figure}

Fits to the combined STAR+PHENIX+BRAHMS data shed further light to the issue of 
short lever arm of the PHENIX and BRAHMS data sets. Namely, we have found out 
that the resulting cross normalization factors $f_P$ and $f_B$ are unrealistically 
small ($\approx 0.8$) if one takes into account the nucleons measured by PHENIX 
and BRAHMS collaborations. Both of these factors can be determined directly from 
the data by dividing the experimental rapidity densities of PHENIX and BRAHMS 
hadrons by the corresponding ones from STAR collaboration. We have plotted the 
ratios of average pions and kaons, i.e. $\pi^++\pi^-$ and $K^++K^-$ from PHENIX 
divided by the corresponding quantities from STAR in fig.~\ref{picrhic}. As one 
can see, the experimental ratios mostly lie between 0.9 and 1.0 at all centralities. 
However, the deviation from unity is clearly large enough so that the previously 
introduced $f_P$ must be implemented. In the same figure, we also show the resulting 
$f_P$ (thick line) fitted to the STAR+PHENIX+BRAHMS data when excluding protons 
and anti-protons from PHENIX and BRAHMS. One can see that the fitted $f_P$ closely 
follows the experimental ratio of PHENIX and STAR pion rapidity densities.
Thus, it seems that the cross-normalization factor for the nucleons would be 
different and much smaller (0.8 or below) than the cross normalization factor for 
the pions and kaons. Unfortunately, this cannot be estimated directly from the 
data, because, unlike for BRAHMS and PHENIX, nucleons from STAR collaboration 
include all the weak decay products of hyperons. It should be mentioned that 
both BRAHMS and PHENIX weak feeding corrections at 200\agev~are based on the 
same PHENIX $\Lambda$ and $\bar{\Lambda}$ measurement at 130\agev. If the 
underlying assumption of $\Lambda/p$ being constant at all centralities at all 
beam energies at RHIC was not correct, then the weak decay corrected proton and 
anti-proton rapidity densities quoted by PHENIX and BRAHMS would be incorrect as 
well, which could partly explain the failure of statistical hadronization model 
fits to the PHENIX and BRAHMS data alone as well as the unrealistically low 
$f_P$ and $f_B$ from fits to the whole data set. In fact, PHENIX (and BRAHMS) 
weak decay correction is based on the 
$\Lambda^{inc}/p^{excl}$=0.89\tpm0.07~\cite{Adcox:2002au} ratio in which
$\Lambda^{inc}$ denotes the inclusive rapidity density of $\Lambda$ including 
weak feeding from $\Xi$'s and $\Omega$ while $p^{excl}=p-0.64\Lambda^{inc}$ is 
the rapidity density of protons from which feeding from inclusive $\Lambda$'s 
has been subtracted. This can be compared with the corresponding ratio at 
200\agev~measured by the STAR collaboration. In order to estimate the inclusive $\Lambda$ 
rapidity density, we sum up all the contributions $\Lambda^{inc}=\Lambda^{excl} 
+ (\Xi^-\rightarrow\Lambda) + (\Omega\rightarrow\Lambda) + (\Xi^0\rightarrow\Lambda)
= 16.7 + 2.17 + 0.68 \times 0.53/2.0 + \oforder{2}$. The $\Xi^0$ rapidity density is 
not measured but is expected to be $\Xi^0\le\Xi^-$. Thus, we get 
$\Lambda^{inc}/p^{excl}\approx 21 / (34.7 - 0.64 \times 21)\approx 1$, clearly 
different from the value obtained by PHENIX collaboration at 130\agev. For comparison, we 
note from the table \ref{input130} that the STAR data at 130\agev~suggests that 
again, $\Lambda^{inc}/p^{excl}\approx 16.2 / (26.4 - 0.64 \times 16.2)\approx 1$. 
Because of this significant discrepancy in the $\Lambda/p$ among the three experiments, we 
have decided to exclude the proton and anti-proton rapidity densities measured by 
PHENIX and BRAHMS collaborations in the analysis and all results in this paper are 
evaluated excluding these 4 measurements. This way, our fitted cross-normalization 
factors $f_P$ and $f_B$ follow the actual ratios of pions and kaons determined from 
the data and provide more reliable estimate compared with fits including the $p$ 
and $\bar{p}$ from PHENIX and BRAHMS in which cases the low $\Lambda/p$ ratio would 
bias the fit towards lower temperatures as well as lower $f_P$ and $f_B$.

We have performed fits to the formula (\ref{basic}) in 7 out of 8 and in 8 out of 11 
most central centrality bins at 130 and 200\agev~respectively. There are two reasons 
why we have refrained from estimating the freeze-out parameters in very peripheral
bins. First, the interpolations of lower multiplicity particles, such as hyperons,
become less and less accurate going to more peripheral collisions and using only 
light mesons and nucleons makes the fit unstable, as has been already discussed. 
Secondly, in extreme peripheral collisions, the role played by exact conservation laws 
(so-called canonical suppression), especially for multi-strange baryons, may become 
important. Yet, it is very difficult to make a definite assessment of this effect 
onto rapidity densities rather than full phase space yields. We remind that 
at top SPS energy, the difference in $\Omega$'s fully integrated yield calculated in 
grand-canonical and S-canonical (enforcing vanishing net strangeness) ensembles
is 32\% and 14\% with $\NP$ 16 and 40 respectively~\cite{Becattini:2005xt}.
Most likely, these figures do not change significantly at RHIC and so it is safer
to use the simple grand-canonical formula (\ref{basic}) only when $\NP \ge 50$.

\section{Comparison with previous analyses}

Up to date several statistical hadronization analyses similar to ours have been 
carried out on RHIC Au-Au collisions. As discussed in the introduction, the majority 
of them formed ratios of rapidity densities {\em a posteriori}, thus possibly 
introducing a bias in the estimation~\cite{Becattini:2007wt}. Also, some of the 
data we have been using was not yet available. It is then useful to compare 
our results with previous ones in order to see how much the different input can 
affect the final result.

In~\cite{Cleymans:2004pp}, different combinations of ratios of rapidity 
densities measured by STAR, PHENIX, PHOBOS and BRAHMS collaborations in Au-Au 
collisions at 130\agev~in 3 centrality bins are analyzed. Depending on the used 
set of ratios of rapidity densities, the authors find somewhat different chemical 
freeze-out conditions. In the case of maximal amount of ratios fitted in the analysis, 
their chemical freeze-out parameters agree with ours within the errors. 

STAR collaboration has determined~\cite{Adams:2003xp,Adams:2006ke} the statistical 
hadronization model parameters in Au-Au collisions at 200\agev~by using different 
combinations of ratios of rapidity densities as a function of centrality. Their 
most recent results are in very good agreement with ours.

The inhomogeneous chemical freeze-out model~\cite{Dumitru:2005hr} takes into 
account possible fluctuations in the temperature and baryon number among 
created clusters in a collision event. The model has been applied to determine the 
temperature and baryon chemical potential in Au-Au collisions at 130 and 
200\agev~at RHIC. In this analysis, ratios of rapidity densities are implemented 
and the authors have found out that the choice of ratios included in the analysis 
can indeed bias the outcome. To try to minimize the bias, particle/antiparticle ratios 
are included along ratios of various particles and negative pions. The central 
values of the distributions of temperature and baryon chemical potential are in 
approximate agreement with our findings. 

In~\cite{Andronic:2005yp} ratios of rapidity densities at RHIC energies have been 
used and the effect of including different sets of rapidity densities as well as 
ratios of them is explored. Admittedly, the authors find a difference in the fit 
outcome whether using rapidity densities or ratios or different set of ratios. 
Among the many quoted results, some are in agreement with ours.

To our knowledge, the only other statistical model analysis~\cite{Rafelski:2004dp}
which used rapidity densities themselves instead of forming ratios includes the 
PHENIX $\pi^\pm$, $K^\pm$ $p$ and $\bar{p}$ as well as two ratios measured by the 
STAR collaboration, and finds somewhat lower temperatures than we do, for the same
version of the statistical model. We deem that this discrepancy owes to the poorer 
data set available when that analysis was carried out.

In conclusion, several groups have analyzed the rapidity densities at RHIC 130 and 
200\agev~and the results of all groups seem to agree rather well. The resulting 
baryon chemical potentials agree very well among all groups and seem to be fairly 
insensitive both to the set of particle species included in the
analysis as well as to the details of the version of the statistical hadronization 
model. Chemical freeze-out temperature, on the other hand, shows larger discrepancies 
and seems to be more sensitive to the input data set, which, as has been emphasized, 
is an effect to be expected in fitting different subsets of ratios, also without 
including correlations. Finally, the values and behaviour of $\gs$ as a function 
of centrality are in very good agreement with previous findings, especially 
with the ones~\cite{Cleymans:2004pp,Adams:2006ke} calculated with the 
THERMUS~\cite{Wheaton:2004qb} package.

On the other hand, we observe a worse fit quality at 200\agev~than generally 
reported by previous analyses. This difference is certainly owing to our updated 
and more accurate data set but, again, to some extent this is possibly related to 
having fitted rapidity densities instead of a subset of ratios.

\section{Summary and conclusions}

In summary, we have analyzed within the statistical hadronization model the 
rapidity densities of various hadron species at mid-rapidity in Au-Au 
collisions at 130 and 200\agev~measured by STAR, PHENIX and BRAHMS collaborations 
and determined the relevant statistical hadronization model best fit parameters. 
This completes our previous analyses at lower center-of-mass energies measured 
at SPS and AGS. 

We have used as input data for the analysis only rapidity densities and not ratios 
formed out of them because of the bias introduced in fitting subset of ratios. Although
a direct comparison is not possible because the data set used in this analysis is
the most up-to-date, our results are in general good agreement with those of previous 
analyses, showing that the effective value of the bias introduced by the actual choice 
of ratios therein was small and most likely within the fit error.

We have found out that the statistical hadronization model, as implemented by the
formula (\ref{basic}) can describe the rapidity densities measured at RHIC relatively 
well, although discrepancies between data and model are visible and larger than 
some other groups using ratios in their analyses have reported. This is also 
reflected in the higher $\chi^2/dof$ value that we find at $\sqrt s_{\NN}=200$ 
GeV compared to that at $\sqrt s_{\NN} = 130$ GeV. Since the relative deviations 
between data and model are comparable at both energies, we conclude that the better 
relative accuracy of measurements at the higher energy has overcome the theoretical
accuracy of the simple formula (\ref{basic}). We interpret this not as a failure 
of the statistical model itself, but an indication that corrections to the simple 
assumptions underlying formula (\ref{basic}) would be necessary, like those 
discussed in Sect.~\ref{discussion}.

A major result of our analysis is the stability of the temperature as a function
of centrality, especially at $\sqrt s_{\NN}=200$ GeV, where all values range from 
about 166 to 171.4 MeV, hence with an overall spread of around 3\%. This confirms
previous findings from STAR collaboration. 

The strangeness under-saturation parameter increases mildly from peripheral to 
central collisions where it almost attains 1. Therefore, RHIC data in peripheral 
collisions demonstrates the phenomenon of phase space under-saturation for mid-rapidity 
yields. This is also observed in the dependence of normalized $\phi$ meson yield 
as a function of centrality \cite{Chen:2008ix}: since the temperature is essentially 
constant, this behaviour can only be parameterized with a $\gs$ varying as a 
function of centrality. The authors have recently proposed \cite{Becattini:2008yn} an 
explanation of strangeness under-saturation in terms of a superposition of NN 
collisions and a completely equilibrated hadronic system originated from the 
central core of the collision, where the Quark Gluon Plasma is formed. Such a 
scenario will be investigated in more detail in future works.

\begin{figure}[!t]
$\begin{array}{c@{\hspace{-0.05cm}}c}
        \epsffile{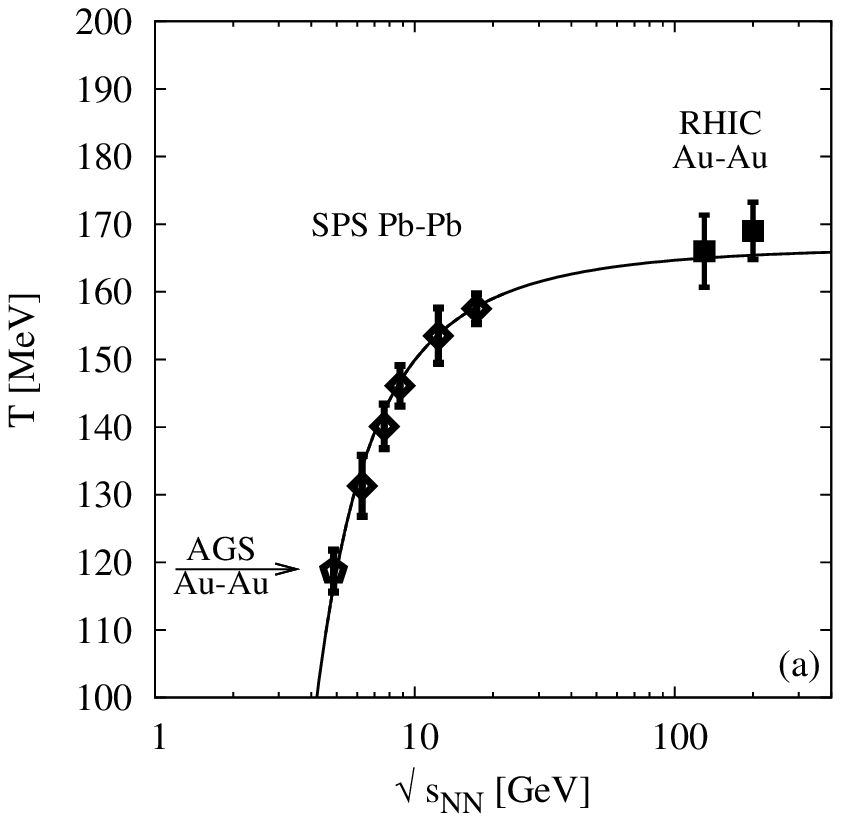} & \epsffile{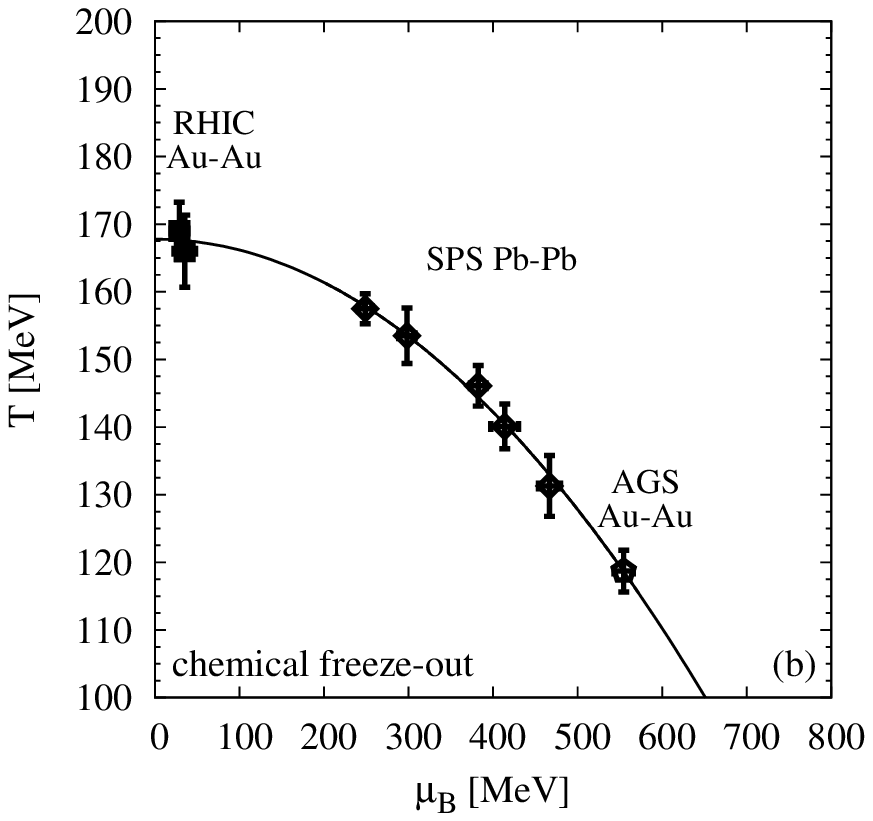} 
\end{array}$
\caption{Chemical freeze-out temperature as a function of center-of-mass
energy per participant pair (left panel (a)) and as a function of baryon chemical 
potential (right panel (b)) in central heavy-ion collisions at AGS, SPS and RHIC. 
The chemical freeze-out shown in the right panel as well as the curve shown 
in the left panel are empirical fits to the AGS and SPS points taken from our 
previous paper~\cite{Becattini:2005xt}.
\label{cfot}}
\end{figure}

Finally, we find that the fitted chemical freeze-out temperatures and baryon 
chemical potentials in central Au-Au collisions nicely fit previously extrapolated 
curves from lower heavy ion collision energies as shown in fig.~\ref{cfot}.

\section*{Acknowledgments}

 We are greatly indebted with H. Caines and J. H. Chen for their help in 
 using STAR data. We acknowledge stimulating and useful discussions with 
 H. Satz. One of the authors (J.M.) was partly supported by the 
 Academy of Finland, project 114371.



\clearpage
\section*{Appendix}
\vspace{-0.5cm}
\begin{table}[!ht]
\begin{center}
\begin{tabular}{|c|c|c|c|c|c|c|c|c|c|}
\hline
& centrality   
& 0-6\%           & 6-11\%        &  11-18\%       & 18-26\%       & 26-34\%       & 34-45\%      & 45-58\%      & 58-85\%\\
& $\frac{\mathrm{d} N_\mathrm{h^-}}{\rm{d}\eta}$~\cite{Adler:2002wn}             
& 290             & 236           & 196            & 154           & 115           & 78.9         & 47.3         & 17.9\\
& $\NP$
&  352           & 279           & 226            & 172           & 126           & 85         & 47         & 18 \\
\hline
&$\pi^+$  &  234\tpm    24  &  190\tpm   19  &  158\tpm   16  &  124\tpm   13  &  92.7\tpm   9.3  &  63.6\tpm   6.4  &  38.1\tpm   3.9 & \\
&$\pi^-$   &  234\tpm   24  &  190\tpm   19  &  158\tpm   16  &  124\tpm   13  &  92.7\tpm   9.3  &  63.6\tpm   6.4  &  38.1\tpm   3.9 & \\
&$p$~\cite{Adams:2003ve}
& \bf  26.37\tpm6.6  & \bf 21.01\tpm5.3 & \bf 16.53\tpm4.1  & \bf 13.03\tpm3.3 & \bf 10.29\tpm2.6 & \bf 7.14\tpm1.8 & \bf 4.36\tpm1.1 & \bf 1.62\tpm0.4  \\
\bf S&$\bar{p}$~\cite{Adams:2003ve}
& \bf 18.72\tpm4.7  & \bf 15.04\tpm3.8 & \bf 11.85\tpm3.0  & \bf 9.50\tpm2.4  & \bf 7.56\tpm1.9  & \bf 5.35\tpm1.3 & \bf 3.31\tpm0.8 & \bf 1.28\tpm0.3  \\
\bf T&$K^+$~\cite{Adler:2002wn}
& \bf  46.2\tpm6.0    & \bf 38.0\tpm4.9   & \bf 28.8\tpm3.8    & \bf 23.1\tpm3.0   & \bf 17.2\tpm2.2  & \bf 11.8\tpm1.5   & \bf 7.23\tpm0.96 & \bf 2.46\tpm0.33  \\
\bf A&$K^-$~\cite{Adler:2002wn}
& \bf 41.9\tpm5.4    & \bf 34.5\tpm4.5   & \bf 26.4\tpm3.5    &\bf 20.8\tpm2.8   &\bf 15.5\tpm2.0  &\bf 10.4\tpm1.4   &\bf 6.48\tpm0.86 &\bf 2.32\tpm0.31  \\
\bf R&$\Lambda$  & 16.2\tpm   1.7  & 12.9\tpm  1.3  &  10.1\tpm  1.0  &  7.34\tpm  0.75  &  5.31\tpm  0.62  &  3.71\tpm  0.82  &  2.23\tpm  0.81 &\\
&$\bar{\Lambda}$ & 11.8\tpm   1.2  & 9.49\tpm  0.94 &  7.52\tpm  0.77 &  5.64\tpm  0.57  &  4.20\tpm  0.46  &  3.01\tpm  0.61  &  1.82\tpm  0.61 &\\
&$\phi$          &  6.26\tpm  0.90 & 5.09\tpm  0.67 &  3.97\tpm  0.56 &  2.77\tpm  0.58  &  1.76\tpm  0.64  &  0.93\tpm  0.62  &  0.43\tpm  0.40 &\\
&$\Xi^-$         &  2.18\tpm  0.28 & 1.74\tpm  0.20 &  1.32\tpm  0.17 &  0.89\tpm  0.18  &  0.53\tpm  0.20  & 0.25\tpm   0.17  & 0.110\tpm 0.059 &\\
&$\Xi^+$         &  1.87\tpm  0.24 & 1.47\tpm  0.17 &  1.09\tpm  0.14 &  0.70\tpm  0.15  &  0.38\tpm  0.17  & 0.15\tpm   0.15  & 0.062\tpm 0.050 &\\
\hline
\bf P&$\pi^+$   & 271\tpm   35  & 228\tpm  29  &  186\tpm   24  &  142\tpm    19  &  104\tpm   14   &  69.7\tpm   9.0  &   39.2\tpm  6.2 &\\
\bf H&$\pi^-$   & 260\tpm   34  & 214\tpm  27  &  171\tpm   22  &  128\tpm    17  &   93\tpm   12   &  63.0\tpm   8.2  &   38.5\tpm  6.0 &\\
\bf E&$K^+$     & 45.4\tpm  6.8 & 37.4\tpm 5.5 &  29.9\tpm  4.4 &  22.2\tpm   3.4 &  15.8\tpm  2.5  &  10.2\tpm   1.6  &   5.5\tpm   1.1 &\\
\bf N&$K^-$     & 40.3\tpm  6.2 & 31.1\tpm 4.7 &  23.1\tpm  3.5 &  15.7\tpm   2.5 &  10.4\tpm  1.7  &   6.8\tpm   1.1  &   4.87\tpm  0.98 &\\
\bf I&$p$       & 28.1\tpm  4.0 & 22.9\tpm 3.1 &  18.1\tpm  2.5 &  13.3\tpm   1.9 &   9.4\tpm  1.4  &   6.05\tpm  0.88 &   3.43\tpm  0.66 &\\
\bf X&$\bar{p}$ & 18.4\tpm  2.7 & 15.0\tpm 2.1 &  11.9\tpm  1.7 &   8.8\tpm   1.3 &   6.29\tpm 0.97 &   4.19\tpm  0.62 &   2.52\tpm  0.48 &\\
\hline
\hline
&centrality   & 0-5\%   & 5-10\%  &  10-20\% & \multicolumn{2}{|c|}{20-35\%}  & \multicolumn{3}{|c|}{35-75\%} \\
& $\frac{\mathrm{d} N_\mathrm{h^-}}{\rm{d}\eta}$~\cite{Adler:2002uv} 
& 296.6   & 243.4   &  186.7   & \multicolumn{2}{|c|}{109.6}    & \multicolumn{3}{|c|}{33.3}   \\    
\hline
&$\Lambda$~\cite{Adler:2002uv}\footnote{Weighted average} 
& \bf 17.2\tpm1.8     & \bf 13.2\tpm1.4               & \bf 10.3\tpm1.0         & \multicolumn{2}{|c|}{\bf 6.0\tpm0.6}      & \multicolumn{3}{|c|}{\bf 1.63\tpm0.17} \\    
&$\bar{\Lambda}$~\cite{Adler:2002uv}$^\mathrm{a}$  
&  \bf 12.2\tpm1.3      & \bf 9.7\tpm1.0              & \bf 7.5\tpm0.8          &\multicolumn{2}{|c|}{\bf 4.7\tpm0.5}      & \multicolumn{3}{|c|}{\bf 1.28\tpm0.13}  \\      
\hline    
&$\pi^+$~\cite{Adams:2003yh}             & \bf 239\tpm11 & \multicolumn{7}{|c|}{}\\    
&$\pi^-$~\cite{Adams:2003yh}             & \bf 239\tpm11 & \multicolumn{7}{|c|}{}\\   
\hline
&centrality   & \mc{2}{0-11\%}       & \mc{2}{11-26\%}       & \mc{4}{26-85\%} \\
\hline
&$\phi$~\cite{Adler:2002xv}
             & \mc{2}{\bf 5.73\tpm0.78} & \mc{2}{\bf 3.33\tpm0.55}  & \mc{4}{\bf 0.98\tpm0.17} \\
\hline
&centrality   & \mc{2}{0-10\%}   & \mc{2}{10-25\%}  &  \mc{4}{25-75\%}   \\
\hline
&$\Xi^-$~\cite{Adams:2003fy}$^\mathrm{a}$ 
& \mc{2}{\bf 2.02\tpm0.25}      & \mc{2}{\bf 1.15\tpm0.17}   & \mc{4}{\bf 0.28\tpm0.04} \\ 
&$\bar{\Xi}^+$~\cite{Adams:2003fy}$^\mathrm{a}$ 
& \mc{2}{\bf 1.72\tpm0.21}      & \mc{2}{\bf 0.93\tpm0.14}    & \mc{4}{\bf 0.22\tpm0.03}   \\
\hline 
&$\Omega+\bar{\Omega}$~\cite{Adams:2003fy}
             &  \mc{2}{\bf 0.56\tpm0.12}  & \multicolumn{6}{|c|}{} \\
\hline
\hline
 &centrality    & 0-5\%              & \mc{2}{5-15\% }            & \mc{2}{15-30\% }           & \mc{2}{30-60\%}            &   60-92\%           \\
&\NPt             & 348\tpm10    & \mc{2}{271.3\tpm8.4}       & \mc{2}{180.2\tpm6.6}       & \mc{2}{78.5\tpm4.6}        & 14.3\tpm3.3         \\
\hline               
\bf P&$\pi^+$~\cite{Adcox:2003nr}    & \bf 276\tpm36   & \mc{2}{\bf 216\tpm28}   & \mc{2}{\bf 141\tpm18}   & \mc{2}{\bf 57.0\tpm7.4} & \bf 9.6\tpm1.2   \\
\bf H&$\pi^-$~\cite{Adcox:2003nr}    & \bf 270\tpm35   & \mc{2}{\bf 200\tpm26}   & \mc{2}{\bf 129\tpm17}   & \mc{2}{\bf 53.3\tpm6.9} & \bf 8.6\tpm1.1    \\
\bf E&$K^+$~\cite{Adcox:2003nr}      & \bf 46.7\tpm7.2 & \mc{2}{\bf 35\tpm5.5}   & \mc{2}{\bf 22.2\tpm3.4} & \mc{2}{\bf 8.3\tpm1.2}  & \bf 0.97\tpm0.19   \\
\bf N&$K^-$~\cite{Adcox:2003nr}      & \bf 40.5\tpm6.5 & \mc{2}{\bf 30.4\tpm4.8} & \mc{2}{\bf 15.5\tpm2.4} & \mc{2}{\bf 6.2\tpm1.0}  & \bf 0.98\tpm0.18    \\
\bf I&$p$~\cite{Adcox:2003nr}        & \bf 28.7\tpm4.1 & \mc{2}{\bf 21.6\tpm3.1} & \mc{2}{\bf 13.2\tpm1.8} & \mc{2}{\bf 5.0\tpm0.7}  & \bf 0.73\tpm0.12   \\
\bf X&$\bar{p}$~\cite{Adcox:2003nr}  & \bf 20.1\tpm3.0 & \mc{2}{\bf 13.8\tpm2.0} & \mc{2}{\bf 9.2\tpm1.4}  & \mc{2}{\bf 3.6\tpm0.5}  & {\bf 0.47\tpm0.08}\\
\hline
&$\Lambda$ ~\cite{Adcox:2002au}      & \bf 17.3\tpm4.4 & \mc{7}{}  \\
&$\bar{\Lambda}$~\cite{Adcox:2002au} & \bf 12.7\tpm3.4 & \mc{7}{}  \\
\hline
\end{tabular}
\caption{Our estimates for the STAR (top panel) and PHENIX (middle panel) rapidity densities of various hadron species in different
centrality windows in Au-Au collisions at 130\agev. STAR $K^\pm$, $p$ and $\bar{p}$ are experimental values while all others are
derived from the measured values shown bold face in the bottom part of the table by interpolation described in Section \ref{sec:interpolation}.
The STAR pion rapidity densities are corrected for weak decays while all other rapidity densities include the weak decay products (if any) 
of weakly decaying resonances.}\label{input130}
\end{center}
\end{table}

\begin{table}
\begin{center}
\begin{tabular}{|c|c|c|c|c|c|c|}
\hline
& centrality  & 0-5\%                & 5-10\%               & 10-15\%            & 10-20\%       & 15-20\%       \\
& \NPt          & 351.4\tpm2.9  & 299.0\tpm3.8   & 253.9\tpm4.3  & 235\tpm9 & 215.3\tpm5.3   \\
\hline
& $\pi^+$~\cite{Adams:2003xp,HC}              & \bf 322.2\tpm 19.2   & \bf 257.0\tpm  15.2  & 210.8\tpm  12.7  &  \bf 193.8\tpm  11.4  & 176.6\tpm    10.7  \\
& $\pi^-$~\cite{Adams:2003xp,HC}              & \bf 327.0\tpm 19.5   & \bf 260.7\tpm  15.4  & 213.7\tpm  12.8  &  \bf 196.1\tpm  11.6  & 178.7\tpm    10.8  \\
&$K^+$~\cite{Adams:2003xp,HC}                 & \bf  51.27\tpm 5.92  & \bf  40.82\tpm  4.25 &  32.9\tpm   3.2  &  \bf  29.97\tpm  2.86 &  27.0\tpm     2.6  \\
&$K^-$~\cite{Adams:2003xp,HC}                 & \bf  49.47\tpm 5.71  & \bf  39.78\tpm  4.15 &  31.7\tpm   3.1  &  \bf  28.74\tpm  2.74 &  25.8\tpm     2.4  \\
\bf S & $p$~\cite{Adams:2003xp,HC}            & \bf  34.70\tpm 4.10  & \bf  28.23\tpm  2.99 &  22.0\tpm   2.2  &  \bf  20.12\tpm  1.94 &  18.3\tpm     1.7  \\
\bf T& $\bar{p}$~\cite{Adams:2003xp,HC}       & \bf  26.70\tpm 3.15  & \bf  21.42\tpm  2.27 &  17.1\tpm   1.7   & \bf  15.69\tpm  1.51 &  14.3\tpm    1.4  \\
\bf A & $\Lambda$~\cite{Adams:2006ke}         & \bf  16.7\tpm  1.1   & 13.55\tpm  0.91      &  11.02\tpm  0.77  & \bf  10.0 \tpm  0.7  &   8.98\tpm  0.64  \\
\bf R & $\bar{\Lambda}$~\cite{Adams:2006ke}   & \bf  12.7\tpm  0.9   & 10.36\tpm  0.69      &   8.47\tpm  0.56  & \bf   7.7 \tpm  0.5  &   6.93\tpm  0.46  \\
& $\Xi^-$~\cite{Adams:2006ke}                 & \bf  2.17\tpm  0.20  &  1.86\tpm  0.14      &   1.55\tpm  0.11  & \bf   1.41\tpm  0.09 &   1.268\tpm 0.074  \\
& $\Xi^+$~\cite{Adams:2006ke}                 & \bf  1.83\tpm  0.21  &  1.52\tpm  0.14      &   1.25\tpm  0.11  & \bf   1.14\tpm  0.09 &   1.027\tpm 0.075  \\
&  $\phi$~\cite{HC,Abelev:2007rw,Chen:2008ix}                & \bf  7.95\tpm  0.74  &  6.81\tpm  0.73      &   5.89\tpm  0.53  & \bf   5.37\tpm  0.51 &   4.82\tpm  0.51  \\
& $\Omega + \bar{\Omega}$~\cite{Adams:2006ke} & \bf  0.53\tpm  0.06  &  0.445\tpm 0.044     &   0.368\tpm 0.035 &                      &   0.299\tpm 0.030 \\
 \hline
\bf P & $\pi^+$~\cite{Adler:2003cb}  &  \bf 286.4\tpm    24.2  &  \bf  239.6\tpm    20.5  &   \bf 204.6\tpm    18.0  &                                          & \bf 173.8\tpm    15.6  \\   
\bf H &$\pi^-$~\cite{Adler:2003cb}  &   \bf 281.8\tpm    22.8  &  \bf  238.9\tpm    19.8  &   \bf 198.2\tpm    16.7  &                                          & \bf 167.4\tpm    14.4   \\   
\bf E&$K^+$~\cite{Adler:2003cb}  &   \bf 48.9\tpm    6.3  &  \bf  40.1\tpm    5.1  &   \bf 33.7\tpm    4.3  &                                          & \bf 27.9\tpm    3.6  \\
\bf N&$K^-$~\cite{Adler:2003cb}  &   \bf 45.7\tpm    5.2  &   \bf 37.8\tpm    4.3  &   \bf 31.1\tpm    3.5  &                                          & \bf 25.8\tpm    2.9   \\
\bf I&$p$~\cite{Adler:2003cb}  &   \bf 18.4\tpm    2.6  & \bf  15.3\tpm    2.1  &  \bf 12.8\tpm    1.8  &                                          & \bf 10.6\tpm    1.5  \\
\bf X&$\bar{p}$~\cite{Adler:2003cb}  & \bf  13.5\tpm    1.8  &  \bf 11.4\tpm    1.5  & \bf   9.5\tpm    1.3  &                                           & \bf 7.9\tpm    1.1   \\
\hline
\bf B& $\pi^+$   &   309.8\tpm  32.3  &   256.8\tpm  25.2  &   213.7\tpm    21.3  & \bf  196.2\tpm  19.7 &   178.8\tpm   18.2  \\
\bf R& $\pi^-$   &   302.6\tpm  31.6  &   253.2\tpm  24.9  &   212.2\tpm    21.2  & \bf  195.3\tpm  19.6 &   178.4\tpm   18.2  \\
\bf A& $K^+$     &   49.8\tpm    5.2  &   40.2\tpm   4.0   &   32.7\tpm    3.3   & \bf   29.8\tpm   3.0 &    26.9\tpm    2.8  \\
\bf H& $K^-$     &   44.7\tpm    4.7  &   37.1\tpm  3.7    &   30.7\tpm    3.1   & \bf    28.1\tpm   2.9 &    25.5\tpm    2.6  \\
\bf M& $p$       &   20.0\tpm    2.1  &   17.2\tpm   1.7   &   14.6\tpm    1.5   & \bf   13.4\tpm   1.4 &    12.2\tpm    1.3  \\
\bf S& $\bar{p}$ &   14.8\tpm    1.6  &   12.6\tpm   1.2   &   10.6\tpm    1.1   & \bf    9.7\tpm   1.0 &     8.8\tpm    0.9  \\
\hline
\hline
& centrality  & \mc{2}{0-10\%}        &\mc{3}{} \\
& $\NP$        & \mc{2}{328\tpm6 }   &\mc{3}{} \\
\hline
\bf B& $\pi^+$~\cite{Arsene:2005mr}   & \mc{2}{\bf 283.3\tpm  28.4} &\mc{3}{} \\
\bf R& $\pi^-$~\cite{Arsene:2005mr}   & \mc{2}{\bf 277.9\tpm  27.9} & \mc{3}{} \\
\bf A& $K^+$~\cite{Arsene:2005mr}     & \mc{2}{\bf 45.0\tpm   4.55} & \mc{3}{} \\
\bf H& $K^-$~\cite{Arsene:2005mr}     & \mc{2}{\bf 40.9\tpm   4.14} & \mc{3}{} \\
\bf M& $p$~\cite{Arsene:2005mr}       & \mc{2}{\bf 18.6\tpm   1.87} & \mc{3}{} \\  
\bf S& $\bar{p}$~\cite{Arsene:2005mr} & \mc{2}{\bf 13.7\tpm   1.38} & \mc{3}{} \\  
\hline 
\end{tabular}
\caption{Rapidity densities of various hadrons in Au-Au collisions at 200\agev~in different centrality windows. 
Bold face numbers denote measured values while numbers written with standard fonts denote our estimates.
Our estimates are interpolated from the experimental values as described in the Sect \ref{sec:interpolation}.
The STAR $p$  and $\bar{p}$ rapidity densities do include weak feeding from (multi)strange hyperons while
PHENIX and BRAHMS $p$ and $\bar{p}$ rapidity densities include feeding from $\Sigma$'s only. 
All other rapidity densities in this table are corrected for the weak feeding (if relevant).}
\label{input200_1}
\end{center}
\end{table}

\begin{table}
\begin{center}
\begin{tabular}{|c|c|c|c|c|c|c|c|}
\hline 
& centrality  & 20-30\%           & 30-40\%           & 40-50\%           & 50-60\%          & 60-70\%     & 70-80\%     \\
& \NPt          & 166.6\tpm5.4  & 114.2\tpm4.4 & 74.4\tpm3.8   & 45.5\tpm3.3   & 25.7\tpm3.8   & 13.4\tpm3.0  \\
\hline
& $\pi^+$~\cite{Adams:2003xp,HC}& \bf 134.93\tpm    7.78  & \bf  89.24\tpm    5.13  & \bf  58.66\tpm    3.35  &  \bf 36.24\tpm    2.07  & \bf  21.07\tpm    1.20  &\\
& $\pi^-$~\cite{Adams:2003xp,HC}&  \bf 136.07\tpm    7.84  & \bf  89.64\tpm    5.16  & \bf  58.85\tpm    3.36  & \bf  36.33\tpm    2.07  & \bf  21.13\tpm    1.20  &\\
& $K^+$~\cite{Adams:2003xp,HC}&   \bf 20.48\tpm    1.77   & \bf  13.61\tpm    1.11  & \bf  8.690\tpm   0.680  & \bf  5.400\tpm   0.410  & \bf  2.980\tpm   0.220  &\\
& $K^-$~\cite{Adams:2003xp,HC}&   \bf 19.68\tpm    1.70  &  \bf 13.18\tpm    1.07  & \bf  8.370\tpm   0.660  &  \bf 5.190\tpm   0.390  & \bf  2.890\tpm   0.220  &\\
\bf S & $p$~\cite{Adams:2003xp,HC}&   \bf 14.39\tpm    1.26  & \bf  9.300\tpm   0.760  &  \bf 6.170\tpm   0.480  &  \bf 3.880\tpm   0.290  &  \bf 2.200\tpm   0.160   &\\
\bf T & $\bar{p}$~\cite{Adams:2003xp,HC}&  \bf 11.180\tpm   0.980  & \bf  7.460\tpm   0.610  &\bf   4.930\tpm   0.390  & \bf  3.160\tpm   0.240  & \bf  1.840\tpm   0.140 &\\
\bf A& $\Lambda$             & 6.67\tpm   0.47   & 4.39\tpm   0.34   & 2.71\tpm   0.21   & 1.42\tpm   0.16      & & \\
\bf R & $\bar{\Lambda}$      & 5.18\tpm   0.35   & 3.43\tpm   0.27   & 2.13\tpm   0.17   & 1.15\tpm   0.12      & & \\
& $\Xi^-$                    & 0.903\tpm  0.037  & 0.537\tpm  0.031  & 0.315\tpm  0.025  & 0.205\tpm  0.035     & & \\
& $\Xi^+$                    & 0.756\tpm  0.043  & 0.484\tpm  0.037  & 0.295\tpm  0.029  & 0.165\tpm  0.031     & & \\
& $\phi$~\cite{HC,Abelev:2007rw,Chen:2008ix}& \bf 3.47\tpm 0.44 & \bf 2.29\tpm 0.23 & \bf 1.44\tpm 0.14 & \bf 0.810\tpm  0.092 & \bf 0.450\tpm  0.051 & \bf 0.20\tpm 0.022 \\
& $\Omega + \bar{\Omega}$    &  0.213\tpm  0.025 & 0.127\tpm  0.020  & 0.073\tpm  0.011  & 0.053\tpm  0.015     & & \\
\hline
\bf P & $\pi^+$~\cite{Adler:2003cb}&  \bf 130.3\tpm    12.4  & \bf  87.0\tpm    8.6  & \bf  54.9\tpm    5.6  & \bf  32.4\tpm    3.4  & \bf  17.0\tpm    1.8 & \bf 7.9\tpm 0.8 \\
\bf H &$\pi^-$~\cite{Adler:2003cb}& \bf 127.3\tpm    11.6  & \bf  84.4\tpm    8.0  & \bf  52.9\tpm    5.2  & \bf  31.3\tpm    3.1  & \bf  16.3\tpm    1.6 &  \bf 7.7\tpm 0.7 \\
\bf E & $K^+$~\cite{Adler:2003cb}&  \bf 20.6\tpm    2.6  &   \bf 13.2\tpm    1.7  & \bf  8.0\tpm   0.8  & \bf  4.5\tpm   0.4  &  \bf 2.2\tpm   0.2 & \bf 0.89\tpm 0.09 \\
\bf N & $K^-$~\cite{Adler:2003cb}&   \bf 19.1\tpm    2.2  &   \bf 12.3\tpm    1.4  &  \bf 7.4\tpm   0.6  & \bf  4.1\tpm   0.4  & \bf  2.0\tpm   0.1 &\bf 0.88\tpm 0.09 \\
\bf I & $p$~\cite{Adler:2003cb}&    \bf 8.1\tpm    1.1  &   \bf 5.3\tpm   0.7  & \bf  3.2\tpm   0.5  & \bf  1.8\tpm   0.3  & \bf  0.93\tpm   0.15 &\bf 0.40\tpm 0.07 \\
\bf X &$\bar{p}$~\cite{Adler:2003cb}&  \bf  5.9\tpm   0.8  & \bf  3.9\tpm   0.5  & \bf  2.4\tpm   0.3  & \bf  1.4\tpm   0.2  & \bf  0.71\tpm   0.12 & \bf 0.29\tpm 0.05 \\
\hline
\bf B &$\pi^+$   &139.2\tpm    13.9  &  100.2\tpm  11.4  &   67.1\tpm    9.3  &   27.9\tpm    7.7  &&\\
\bf R &$\pi^-$   &138.9\tpm    13.9  &   98.7\tpm  11.2  &   64.0\tpm    9.1  &   27.7\tpm    7.7  &  &\\
\bf A &$K^+$     & 20.8\tpm    2.1  &    15.1\tpm  1.8  &    10.0\tpm    1.5  &   2.7\tpm   1.3  & &\\
\bf H &$K^-$     & 19.4\tpm    2.0  &    13.3\tpm  1.6  &    8.4\tpm     1.4  &   3.4\tpm   1.2  &&\\
\bf M &$p$       &  9.22\tpm   0.93  &   5.98\tpm  0.71  &   3.61\tpm    0.58 &   2.02\tpm   0.54  & &\\
\bf S & $\bar{p}$&  6.70\tpm   0.68  &   4.50\tpm  0.53  &   2.86\tpm    0.46 &   1.58\tpm   0.42  &&\\
\hline
\hline
& centrality   &  \mc{2}{20-40\% }  & \mc{2}{40-60\%}  & \mc{2}{60-80\% }   \\
& $\NP$        & \mc{2}{141\tpm8} & \mc{2}{62\tpm9}   & \mc{2}{21\tpm6}  \\
\hline
\bf S & $\Lambda$~\cite{Adams:2006ke}         & \mc{2}{\bf 5.53\tpm 0.39} & \mc{2}{\bf 2.07\tpm  0.14}  & \mc{2}{\bf 0.58\tpm 0.04}  \\
\bf T & $\bar{\Lambda}$~\cite{Adams:2006ke}   & \mc{2}{\bf 4.30\tpm 0.30} & \mc{2}{\bf 1.64\tpm  0.11}  & \mc{2}{\bf 0.48\tpm 0.03}  \\
\bf A & $\Xi^-$~\cite{Adams:2006ke}           & \mc{2}{\bf 0.72\tpm 0.03} & \mc{2}{\bf 0.26\tpm  0.02}  & \mc{2}{\bf 0.063\tpm 0.005} \\
\bf R & $\bar{\Xi}^+$~\cite{Adams:2006ke}     & \mc{2}{\bf 0.62\tpm 0.04} & \mc{2}{\bf 0.23\tpm  0.02}  & \mc{2}{\bf 0.061\tpm 0.004}  \\
& $\Omega + \bar{\Omega}$~\cite{Adams:2006ke} & \mc{2}{\bf 0.17\tpm 0.02} & \mc{2}{\bf 0.063\tpm 0.009} & \mc{2}{}  \\
\hline
\bf B&$\pi^+$~\cite{Arsene:2005mr}     & \mc{2}{\bf 119.7\tpm  12.1}  & \mc{2}{\bf   47.5\tpm   4.85} & \mc{2}{}\\
\bf R&$\pi^-$~\cite{Arsene:2005mr}     & \mc{2}{\bf 118.8\tpm  12.0}  & \mc{2}{ \bf  46.3\tpm   4.71} & \mc{2}{}\\
\bf A&$K^+$~\cite{Arsene:2005mr}       & \mc{2}{\bf  17.9\tpm   1.83} & \mc{2}{\bf    6.3\tpm   0.68} & \mc{2}{}\\
\bf H&$K^-$~\cite{Arsene:2005mr}       & \mc{2}{\bf  16.3\tpm   1.7}  & \mc{2}{\bf    5.9\tpm   0.64} & \mc{2}{}\\
\bf M&p~\cite{Arsene:2005mr}           & \mc{2}{\bf   7.6\tpm   0.77} & \mc{2}{\bf    2.81\tpm   0.29}& \mc{2}{}\\
\bf S&$\bar{p}$~\cite{Arsene:2005mr}   & \mc{2}{\bf   5.6\tpm   0.6}  & \mc{2}{\bf    2.22\tpm   0.24}& \mc{2}{}\\
\hline
\end{tabular}
\caption{Rapidity densities of various hadrons in Au-Au collisions at 200\agev~in different centrality windows. 
Bold face numbers denote measured values while numbers written with standard fonts denote our estimates.
Our estimates are interpolated from the experimental values as described in the Sect \ref{sec:interpolation}.
The weak decay corrections are the same as listed in the previous table.} 
\label{input200_2}
\end{center}
\end{table}

\begin{table}[!ht]
\begin{center}
\begin{tabular}{|c|c|c|c|c|c|c|}
\hline
centrality  & $T$ [MeV] & $\mu_B$ [MeV] & $\gamma_S$ & $\d V/\d y$ [fm$^3$] & $\chi^2$/dof & $f_P$\\
\hline 
& \mc{6}{STAR 130\agev~best fit parameters} \\
\hline
0-6\%        &  165.9\tpm   5.3 & 35.1\tpm  12.6 & 1.109\tpm  0.078 & 1097\tpm   258 &    5.9 /   7 &  \\ 
6-11\%       &  165.3\tpm   5.1 & 34.0\tpm  12.2 & 1.104\tpm  0.075 &  925\tpm   211 &    5.4 /   7 &  \\ 
11-18\%      &  165.3\tpm   5.3 & 33.2\tpm  12.6 & 1.053\tpm  0.075 &  760\tpm   176 &    6.1 /   7 &  \\
18-26\%      &  162.5\tpm   5.4 & 30.7\tpm  13.3 & 0.977\tpm  0.078 &  712\tpm   173 &    2.6 /   7 &  \\
26-34\%      &  163.1\tpm   2.5 & 28.1\tpm  15.0 & 0.907\tpm  0.056 &  534\tpm   61  &    1.4 /   7 &  \\
34-45\%      &  161.1\tpm   7.4 & 27.3\tpm  21.8 & 0.863\tpm  0.091 &  410\tpm   140 &    2.5 /   7 &  \\
45-58\%      &  153.5\tpm   7.8 & 26.9\tpm  26.1 & 0.823\tpm  0.096 &  352\tpm   136 &    4.7 /   7 &  \\
\hline
& \mc{6}{PHENIX 130\agev~best fit parameters} \\
\hline
 0-5\%             &   161.4\tpm     6.9  & 33.9\tpm  16.8 & 1.02\tpm 0.16   & 1397\tpm   475 &    1.7 /   2   &  \\ 
\hline
& \mc{6}{STAR+PHENIX 130\agev~best fit parameters} \\
\hline
0-6\%        &  163.8 \tpm   4.1  & 36.3 \tpm  10.2   & 1.109 \tpm  0.067    & 1225 \tpm  228  &    8.2 /  12  &  0.919\tpm 0.067 \\
6-11\%       &  163.7 \tpm   4.0  & 36.1 \tpm  9.91   & 1.097 \tpm  0.064    & 1013 \tpm  182  &    8.6 /  12  &  0.911\tpm 0.065 \\ 
11-18\%      &  163.8 \tpm   4.0  & 35.9 \tpm  10.1   & 1.043 \tpm  0.064    &  833 \tpm  150  &    10.3 /  12 &  0.897\tpm 0.065 \\
18-26\%      &  161.9 \tpm   3.9  & 34.4 \tpm  10.7   & 0.954 \tpm  0.024    &  750 \tpm  133  &    7.2 /  12  &  0.844\tpm 0.060 \\
26-34\%      &  162.1 \tpm   4.4  & 32.9 \tpm  5.26   & 0.884 \tpm  0.067    &  576 \tpm  116  &    7.1 /  12  &  0.801\tpm 0.061 \\
34-45\%      &  159.1 \tpm   4.9  & 32.7 \tpm  13.6   & 0.827 \tpm  0.071    &  460 \tpm  106  &    7.5 /  12  &  0.798\tpm 0.064 \\
45-58\%      &  153.3 \tpm   5.4  & 26.8 \tpm  18.3   & 0.813 \tpm  0.080    &  357 \tpm   96  &    5.0 /  12  &  0.841\tpm 0.018 \\ 
\hline
\end{tabular}
\caption{Statistical hadronization model best fit parameters at chemical freeze-out in Au-Au collisions at 130\agev.}
\label{parameters130}
\end{center}
\end{table}

\begin{table}[!hT]
\begin{center}
\begin{tabular}{|c|c|c|c|c|c|c|c|}
\hline
centrality  & $T$ [MeV] & $\mu_B$ [MeV] & $\gamma_S$ & $\d V/\d y$ [fm$^3$] & $\chi^2$/dof & $f_P$ & $f_B$ \\
\hline 
& \mc{7}{STAR 200\agev~best fit parameters} \\
\hline
0-5\%   &     168.0\tpm    6.2&      28.8\tpm      14.3 &       0.935\tpm      0.064 &       1419\tpm       377 &       22.2  /  8 & & \\
5-10\%  &     169.5\tpm    6.8&      29.1\tpm      14.7 &       0.941\tpm      0.069 &       1055\tpm       304 &       26.8  /  8 & & \\
10-15\% &     167.0\tpm    7.3&      26.8\tpm      16.2 &       0.979\tpm      0.079 &        941\tpm       296 &       36.9  /  8 & & \\
10-20\% &     168.8\tpm    6.0&      27.0\tpm      13.3 &       1.054\tpm      0.077 &        745\tpm       190 &       22.4  /  7 & & \\
15-20\% &     167.8\tpm    7.0&      27.1\tpm      15.3 &       0.971\tpm      0.076 &        750\tpm       225 &       36.1  /  8 & & \\
20-30\% &     169.2\tpm    6.3&      27.9\tpm      12.5 &       0.954\tpm      0.064 &        537\tpm       141 &       30.0  /  8 & & \\
30-40\% &     166.4\tpm    5.5&      22.0\tpm      12.7 &       0.951\tpm      0.063 &        399\tpm        95 &       26.7  /  8 & & \\
40-50\% &     165.8\tpm    5.2&      21.4\tpm      12.8 &       0.900\tpm      0.059 &        274\tpm        62 &       25.1  /  8 & & \\
50-60\% &     164.9\tpm    3.1&      20.5\tpm       8.0 &       0.902\tpm      0.043 &        173\tpm        24 &        7.0  /  8 & & \\
\hline
& \mc{7}{STAR+PHENIX+BRAHMS 200\agev~best fit parameters} \\
\hline
0-5\%   &     169.2\tpm      5.2&      29.5\tpm      11.2 &       0.929\tpm      0.044&       1336\tpm       302 &       23.4  / 14 & 0.863\tpm 0.073& 0.89\tpm 0.12\\
5-10\%  &     171.2\tpm      5.2&      29.7\tpm      11.3 &       0.928\tpm      0.048&        976\tpm       210 &       28.1  / 14 & 0.912\tpm 0.078& 0.931\tpm 0.082 \\
10-15\% &     168.9\tpm      5.5&      27.6\tpm      12.6 &       0.960\tpm      0.054&        868\tpm       201 &       39.7  / 14 & 0.928\tpm 0.093& 0.932\tpm 0.096 \\
15-20\% &     169.9\tpm      5.4&      27.9\tpm      12.0 &       0.951\tpm      0.051&        686\tpm       156 &       38.9  / 14 & 0.948\tpm 0.095& 0.944\tpm 0.098 \\
20-30\% &     171.4\tpm      5.0&      28.8\tpm       9.90&       0.935\tpm      0.046&        489\tpm       102 &       32.8  / 14 & 0.942\tpm 0.088& 0.970\tpm 0.091 \\
30-40\% &     168.3\tpm      4.4&      23.1\tpm      10.2 &       0.930\tpm      0.043&        367\tpm        68 &       30.0  / 14 & 0.931\tpm 0.085 & 1.04\tpm 0.10 \\
40-50\% &     167.3\tpm      3.1&      22.5\tpm       9.35&       0.883\tpm      0.030&        257\tpm       34&       27.9  / 14 & 0.870\tpm 0.068& 1.06\tpm 0.12 \\
50-60\% &     166.2\tpm      3.3&      20.8\tpm      10.4 &       0.874\tpm      0.045 &       165\tpm       24&       11.1  / 14 & 0.818\tpm 0.051& 0.67\tpm0.11\\
\hline
\end{tabular}
\caption{Statistical hadronization model best fit parameters at chemical freeze-out in Au-Au collisions at 200\agev. 
Errors are scaled according to the PDG scaling scheme~\cite{pdg} by a factor $\sqrt{\chi^2 / dof}$.}
\label{parameters200}
\end{center}
\end{table}

\clearpage

\begin{table}[!ht]\begin{center}
\begin{tabular}{|c|c|c|c|c|}
\hline
        &        Experiment (E)       & Model (M)  & Residual & (M - E)/E  (\%)\\ 
\hline
& \mc{4}{STAR Au-Au 130\agev~0-6\% most central collisions} \\
\hline
$\pi^+$            &       234\tpm       24    &        213   &       -0.87     &      -8.93    \\ 
$\pi^-$            &       234\tpm       24    &        215   &       -0.76     &      -7.82    \\ 
$K^+$              &       46.2\tpm       6.0  &        49.0  &        0.46     &       5.96    \\ 
$K^-$              &       41.9\tpm       5.4  &        45.7  &        0.69     &       8.95    \\ 
$p$                &       26.4\tpm       5.8  &        32.2  &         1.0     &       22.3    \\ 
$\bar{p}$          &       18.7\tpm       4.1  &        22.4  &        0.90     &       19.8    \\ 
$\phi$             &       6.26\tpm      0.90  &        6.61  &        0.38     &       5.48    \\ 
$\Lambda$          &       16.2\tpm       1.6  &        16.3  &       0.064     &      0.648    \\ 
$\bar{\Lambda}$    &       11.8\tpm       1.2  &        12.2  &        0.32     &       3.28    \\ 
$\Xi^-$            &       2.18\tpm      0.28  &        2.00  &       -0.63     &      -8.24    \\ 
$\bar{\Xi}^+$      &       1.87\tpm      0.24  &        1.59  &        -1.2     &      -14.9    \\ 
\hline
\end{tabular}\end{center}
\caption{Comparison between estimated rapidity densities in the combined fit
and rapidity densities measured by STAR in central Au-Au collisions at 130\agev. 
The 3rd and 4th column show the discrepancy between data and model in units of 
standard error and in percentages, respectively.}
\label{combo130_ff1}
\end{table}

\begin{table}[!ht]\begin{center}
\begin{tabular}{|c|c|c|c|c|}
\hline
        &        Experiment (E)       & Model (M)  & Residual & (M - E)/E  (\%)\\ 
\hline
& \mc{4}{STAR Au-Au 200\agev~0-5\%  most central collisions} \\
\hline
$\pi^+$            &       322\tpm       19    &        326   &        0.21     &       1.24    \\ 
$\pi^-$            &       327\tpm       20    &        328   &       0.055     &      0.326    \\ 
$K^+$              &       51.3\tpm      5.9   &        57.1  &        0.99     &       11.4    \\ 
$K^-$              &       49.5\tpm       5.7  &        53.9  &        0.78     &       8.97    \\ 
$p$                &       34.7\tpm       4.1  &        41.8  &         1.7     &       20.4    \\ 
$\bar{p}$          &       26.7\tpm       3.1  &        30.9  &         1.3     &       15.9    \\ 
$\phi$             &       7.95\tpm      0.74  &        6.73  &        -1.6     &      -15.3    \\ 
$\Lambda$          &       16.7\tpm       1.1  &        14.4  &        -2.1     &      -13.9    \\ 
$\bar{\Lambda}$    &      12.70\tpm      0.92  &       11.07  &        -1.8     &      -12.8    \\ 
$\Xi^-$            &       2.17\tpm      0.20  &        2.02  &       -0.74     &      -6.86    \\ 
$\bar{\Xi}^+$      &       1.83\tpm      0.21  &        1.67  &       -0.76     &      -8.51    \\ 
$\Omega+\bar{\Omega}$&    0.530\tpm     0.057  &       0.651  &         2.1     &       22.9    \\ 
\hline
\end{tabular}\end{center}
\caption{Comparison between estimated rapidity densities in the combined fit
and rapidity densities measured by STAR in central Au-Au collisions at 200\agev. 
The 3rd and 4th column show the discrepancy between data and model in units of 
standard error and in percentages, respectively.}
\label{combo200_f1}
\end{table}

\end{document}